\documentclass[journal]{IEEEtran}

\usepackage{cite}

\usepackage[bookmarks, colorlinks=true, plainpages = false, citecolor = blue,linkcolor=red,urlcolor = blue, filecolor = blue,pagebackref]{hyperref}

\usepackage{setspace,multicol,color}

\usepackage{graphicx,graphics}
\usepackage{subfig,epsfig}

\usepackage{bm,amsmath,amssymb,amsthm,enumerate}

\usepackage{algorithmicx}
\usepackage[linesnumberedhidden,ruled,vlined]{algorithm2e}

\usepackage{wrapfig}

\usepackage{tabularx}

\theoremstyle{plain}
\newtheorem{lemma}{\textbf{Lemma}}
\newtheorem{theorem}{\textbf{Theorem}}
\newtheorem{corollary}{\textbf{Corollary}}

\theoremstyle{definition}
\theoremstyle{remark}

\newcommand{\bLambda}{{\bm{\Lambda}}}

\newcommand{\cA}{{\mathcal{A}}}
\newcommand{\cB}{{\mathcal{B}}}
\newcommand{\cF}{{\mathcal{F}}}
\newcommand{\cP}{{\mathcal{P}}}

\newcommand{\cS}{{\mathcal{S}}}

\newcommand{\cR}{{\mathcal{R}}}
\newcommand{\cH}{{\mathcal{H}}}

\newcommand{\cN}{{\mathcal{N}}}
\newcommand{\cM}{{\mathcal{M}}}

\newcommand{\bb}{{\boldsymbol b}}

\newcommand{\be}{{\boldsymbol e}}

\newcommand{\bt}{{\boldsymbol t}}
\newcommand{\bu}{{\boldsymbol u}}

\newcommand{\bw}{{\boldsymbol w}}
\newcommand{\bx}{{\boldsymbol x}}
\newcommand{\by}{{\boldsymbol y}}

\newcommand{\E}{{\mathbb E}}
\newcommand{\R}{{\mathbb R}}

\newcommand{\bA}{{\boldsymbol A}}
\newcommand{\bB}{{\boldsymbol B}}
\newcommand{\bC}{{\boldsymbol C}}
\newcommand{\bD}{{\boldsymbol D}}

\newcommand{\bF}{{\boldsymbol F}}

\newcommand{\bH}{{\boldsymbol H}}
\newcommand{\bI}{{\boldsymbol I}}

\newcommand{\bK}{{\boldsymbol K}}
\newcommand{\bL}{{\boldsymbol L}}

\newcommand{\bN}{{\boldsymbol N}}

\newcommand{\bP}{{\boldsymbol P}}
\newcommand{\bQ}{{\boldsymbol Q}}
\newcommand{\bR}{{\boldsymbol R}}
\newcommand{\bS}{{\boldsymbol S}}

\newcommand{\bU}{{\boldsymbol U}}

\newcommand{\bX}{{\boldsymbol X}}

\newcommand{\bZ}{{\boldsymbol Z}}

\newcommand{\Tr}{\mathop{\rm Tr}}

\newcommand{\argmin}{\mathop{\rm argmin}}

\newcommand{\sgn}{\mathop{\rm sgn}}
\newcommand{\vect}{\mathop{\rm vec}}
\newcommand{\grad}{{\triangledown}}

\renewcommand{\Re}{\operatorname{Re}}

\begin{document}
%
\title{Iterative Thresholding and Projection Algorithms and Model-Based Deep Neural Networks for Sparse LQR Control Design}
%
%
%

\author{Myung~Cho
\thanks{M. Cho is with the Department of Electrical and Computer Engineering, California State University, Northridge, CA, 91330, USA
        {\tt\small (email: michael.cho@csun.edu)}}
\thanks{\textcolor[rgb]{0,0,0}{A part of this journal paper was presented at the European Control Conference (ECC) in 2022 \cite{cho2022iterative}.}}
}

%
%

\markboth{}%
{Shell \MakeLowercase{\textit{et al.}}: Bare Demo of IEEEtran.cls for IEEE Journals}
%



\maketitle

\begin{abstract}
In this paper, we consider an LQR design problem for distributed control systems. For large-scale distributed systems, finding a solution might be computationally demanding due to communications among agents. To this aim, we deal with LQR minimization problem with a regularization for sparse feedback matrix, which can lead to achieve the reduction of the communication links in the distributed control systems. For this work, we introduce simple but efficient iterative algorithms  - Iterative Shrinkage Thresholding Algorithm (ISTA) and Iterative Sparse Projection Algorithm (ISPA). They can give us a trade-off solution between LQR cost and sparsity level on  feedback matrix. Moreover, in order to improve the speed of the proposed algorithms, we design deep neural network models based on the proposed iterative algorithms. Numerical experiments demonstrate that our algorithms can outperform the previous methods using the Alternating Direction Method of Multiplier (ADMM) \cite{lin2013design} and the Gradient Support Pursuit (GraSP) \cite{lian2017game}, and their deep neural network models can improve the performance of the proposed algorithms in convergence speed.
\end{abstract}


%
\IEEEpeerreviewmaketitle

\section{Introduction}
\label{sec:intro}
In control field, optimal control design targets designing a feedback controller minimizing a certain objective function, e.g., Linear Quadratic Regulator (LQR), $\cH_2$ norm or $\cH_{\infty}$ norm. Various methods to design feedback controllers have been studied in the past several decades \cite{rautert1997computational,zhou1996robust,peres1994alternate}. Unlike the system models considered in the previous research \cite{rautert1997computational,zhou1996robust,peres1994alternate}, current control systems can become larger and more complicated by having multi-agents in a distributed manner, which raises different problems and issues from the previous ones such as how to efficiently control large-scale distributed control systems, how to efficiently communicate among the multi-agents in distributed systems, and how to preserve security and privacy issues in distributed systems. 

Besides minimizing LQR, $\cH_2$ norm, or $\cH_{\infty}$ norm for optimal control, in large-scale distributed control systems such as smart grid and multi-agent drone systems, communication among the multi-agents is desired to efficiently control distributed systems. However, due to network constraints such as limited communication bandwidth, limited communication power, and limited response time, the communication among the multi-agents can be a big burden in large-scale distributed control systems. Therefore, it is natural to ask about how to address both the minimization of  LQR, $\cH_2$ norm, or $\cH_{\infty}$ norm and the network constraints on communication simultaneously for the control of large-scale distributed systems with reduced communication burdens.

Designing optimal feedback controllers for distributed and interconnected systems has been studied in various research \cite{bamieh2002distributed,d2003distributed,bamieh2005convex,motee2008optimal,lin2013design, fardad2009optimal,lian2017game,cho2020communication,cho2023low} and the references therein. Among them, we are interested in the optimal control design problem with sparse network constraints considered in \cite{lin2013design, fardad2009optimal,lian2017game} in order to reduce the number of communication links in large-scale distributed control systems.  Since in large-scale distributed control systems,  non-zero elements in off-diagonal of a feedback matrix are related to communication links among multi-agents in a distributed control system, a sparse feedback matrix can be understood as the reduction of communication links in a distributed control system. Thus, designing a sparse feedback matrix by considering network constraints can come into play for the reduction of communication burdens in control of large-scale distributed systems.

For sparse communication links in a distributed control system, the authors in \cite{lin2013design} considered the sum of LQR and a regularization term for sparse-promoting operation on a feedback controller as an objective function in a minimization problem, so-called the sparse LQR control design problem, and designed an algorithm based on the ADMM to solve the optimal control design problem. In \cite{fardad2009optimal}, under the assumption that the structured network model is known, the authors took into account the optimal control design problem. Additionally, the authors in \cite{lian2017game} tackled the optimal control design problem with  sparse network constraints by using the Gradient Support Pursuit (GraSP) \cite{bahmani2013greedy}, which can be understood as an iteratively updating $\ell_0$-ball projection algorithm. In this paper, we take into acount the sparse LQR control design problems considered in \cite{lin2013design, lian2017game} and introduce efficient algorithms and their deep neural network models to solve the problems.

The contribution of this paper is three-fold. Firstly, we introduce simple but efficient algorithms - Iterative Shrinkage-Thresholding Algorithm (ISTA) and Iterative Sparse Projection Algorithm (ISPA) - for the sparse optimal control design problems. Inspired by the well-known iterative thresholding algorithm in the field of signal processing\cite{beck2009fast,daubechies2004iterative,tibshirani2013lasso}, we design the ISTA for the sparse optimal control design problem with handling a feasible set in feedback controller design. In the context of optimization, the major difference between the conventional ISTA and the proposed ISTA is that the optimal control problem has a feasible set of possible feedback matrices stabilizing a control system, while the conventional ISTA deals with the problem with no constraints. For the ISPA, we consider using the projection algorithm for the sparsity-promoting operation. In the numerical experiments, we show that our ISTA and ISPA can provide a trade-off solution between LQR cost and sparsity level on feedback matrices, and our proposed methods can outperform the ADMM-based algorithm introduced in \cite{lin2013design} and the GraSP based method introduced in \cite{lian2017game}, Secondly, we provide the analyses of our algorithms for the sparse LQR control design problems. Finally, we further design Deep Neural Network (DNN) models based on our ISTA and Fast ISTA (FISTA). The DNNs based the ISTA and FISTA can also be understood as the model-based DNN models for the ISPA as well. Numerical experiments demonstrate that the neural network models can further enhance the performance of the proposed algorithms for sparse optimal control design. \textcolor[rgb]{0,0,0}{In the previous conference paper that served as a precursor to this journal paper, we introduced the concept of ISTA for sparse optimal control design, along with its model-based DNN implementation. Building upon the conference precursor, this paper presents significant extensions. Firstly, we introduce a novel algorithm called ISPA. Additionally, we provide comprehensive analyses of the ISTA algorithm and its enhanced DNN counterpart, DNN-FISTA, which is based on the fast ISTA. Furthermore, we include additional simulations involving a power system model to further validate our proposed methods, and a comparison between GraSP \cite{lian2017game} and ISPA.}

The rest of the paper is organized as follows. In Section \ref{sec:signal}, we introduce the problem formulation for the optimal control design problems minimizing LQR with sparsity constraint on communication network. In Section \ref{sec:review}, we review the previous research on sparse LQR control design problems. In Section \ref{sec:algo}, we briefly introduce the ISTA for sparse signal recovery and describe the ISTA for the sparse LQR control design problem. Section \ref{sec:ISPA} describes the ISPA and its relation with the ISTA. Section \ref{sec:DNN} introduces the model-based DNNs based on our proposed algorithms. In Section \ref{sec:experiment}, we  demonstrate the performance of our proposed methods including the ISTA, the ISPA, and the DNN models against the previous research using ADMM \cite{lin2013design} and GraSP \cite{lian2017game}. Finally, we summarize the paper in Section \ref{sec:conclusion}.

\begin{table}[h] 
\caption{Notations}
\small
\begin{tabularx}{0.49\textwidth}{p{0.115\textwidth}|l}
\hline\hline
\hspace{-1em} Non-bold letter               &   Scalar 	    (e.g., $x$, $X$)\\  \hline
\hspace{-1em} Bold small letter             &   Vector 	    (e.g., $\bx$)\\  \hline
\hspace{-1em} Bold captial letter             &   Matrix 	(e.g., $\bX$)   \\  \hline
\hspace{-1em} $\R$               &   The set of real numbers 	    \\  \hline
\hspace{-1em} $\Re(\cdot)$               &   The real part of a complex value	    \\  \hline
\hspace{-1em} Super-script $T$          &   Transpose	    \\  \hline
\hspace{-1em} Super-script $(t)$          &   The $t$-th iteration value, (e.g., $\bK^{(t)}$)	    \\  \hline
\hspace{-1em} $\| \cdot \|_F$        &   Frobenius norm	  \\  \hline
\hspace{-1em} $\| \cdot \|_2$        &   $\ell_2$ norm	  \\  \hline
\hspace{-1em} $\| \cdot \|_1$        &   Element-wise $\ell_1$ norm \\ 
														 &  (e.g., $\|\bK\|_1=\sum_{i,j} |K_{i,j}|$) \\ \hline
\hspace{-1em} $\| \cdot \|_0$        &   Number of non-zero elements \\ \hline
\hspace{-1em} $\bI$        &   Identity matrix \\ \hline
\hspace{-1em} $\lambda(\bX)$        &  Eigenvalue of a matrix $\bX$\\ \hline
\hspace{-1em} $\cF$       &  A feasible set of feedback matrices $\bK$'s\\
										 &  with asymptotic stability,\\ 
										 &  i.e.,  $\cF := \{  \bK \;|\;  \max( \Re( \lambda ( \bA - \bB\bK ) ) ) < 0 \}$\\ \hline
\hspace{-1em} $\bQ \succeq 0$         &  Symmetric positive semidefinite matrix $\bQ$ \\ \hline
\hspace{-1em} $\bQ \succ 0$         &  Symmetric positive definite matrix $\bQ$ \\ \hline
\hspace{-1em} $\otimes$       &  Kronecker product\\ \hline
\hspace{-1em} $\langle \cdot, \cdot \rangle$       &  Inner product\\ \hline
\hspace{-1em} $\max( \cdots )$       &  Maximum value\\ \hline
\hspace{-1em} $\vect(\bK)$       &  Vectorization of a matrix $\bK$ \\
												 	& by stacking the column vectors of $\bK$\\ \hline
\hspace{-1em} $\vect(\bK)_{(i)}$       &  The $i$-th sorted element in a descending order \\
														& of magnitude,  i.e., $|\vect(\bK)_{(1)} | \geq |\vect(\bK)_{(2)} | \; ... $\\ \hline
\hspace{-1em} $\cN(0, \bI)$       & Gaussian normal distribution \\
													& with zero mean and identity co-variance matrix\\ \hline
\hline
\end{tabularx}
\end{table}


\section{Problem Formulation}
\label{sec:signal}
We consider a distributed control system having a closed-loop expressed as the following state space representation:
\par\noindent\small
\begin{align}\label{eq:SSE}
	& \dot{\bx}(t) = \bA \bx(t) + \bB \bu(t),\nonumber  \\
	& \by(t) = \bC \bx(t) + \bD \bu(t), \nonumber \\
	& \bu(t) = -\bK \bx(t),
\end{align}
\normalsize
where  $\bx(t)$ and $\dot{\bx}(t)$ are the state of the system and its derivative at time $t$ respectively organized by stacking the states of agents in the distributed system, $\bu(t)$ is the system input at time $t$ which is also organized by stacking the inputs of agents in the system, and $\by(t)$ is the output of the system at time $t$. Correspondingly $\bA \in \R^{n \times n}$ is a state matrix, $\bB \in \R^{n \times m}$ is an input matrix for system input, $\bC = [\bQ^{1/2} \; \bm{0} ]^T \in \R^{p \times n}$ and $\bD = [ \bm{0} \; \bR^{1/2}]^T \in \R^{p \times m}$ are output matrices, and $\bK \in \R^{m \times n}$ is a feedback matrix. Through out the paper, we assume that $(\bA, \bB)$ is stabilizable and $(\bA, \bQ^{1/2})$ is detectable. Then, we can rewrite \eqref{eq:SSE} as 
\par\noindent\small
\begin{align}\label{eq:SSE_feedback}
	& \dot{\bx}(t) = (\bA - \bB\bK)\bx(t), \nonumber  \\
	& \by(t) = (\bC - \bD\bK) \bx(t).
\end{align}
\normalsize
In this feedback system, we are interested in finding a sparse feedback matrix $\bK$ asymptotically stabilizing the system, i.e., $\bK \in \cF$, and minimizing the energy of transfer function.

\begin{figure}[t]
    \centering
    \subfloat[Illustration of a distributed system having multi-agents]{\includegraphics[scale=0.32]{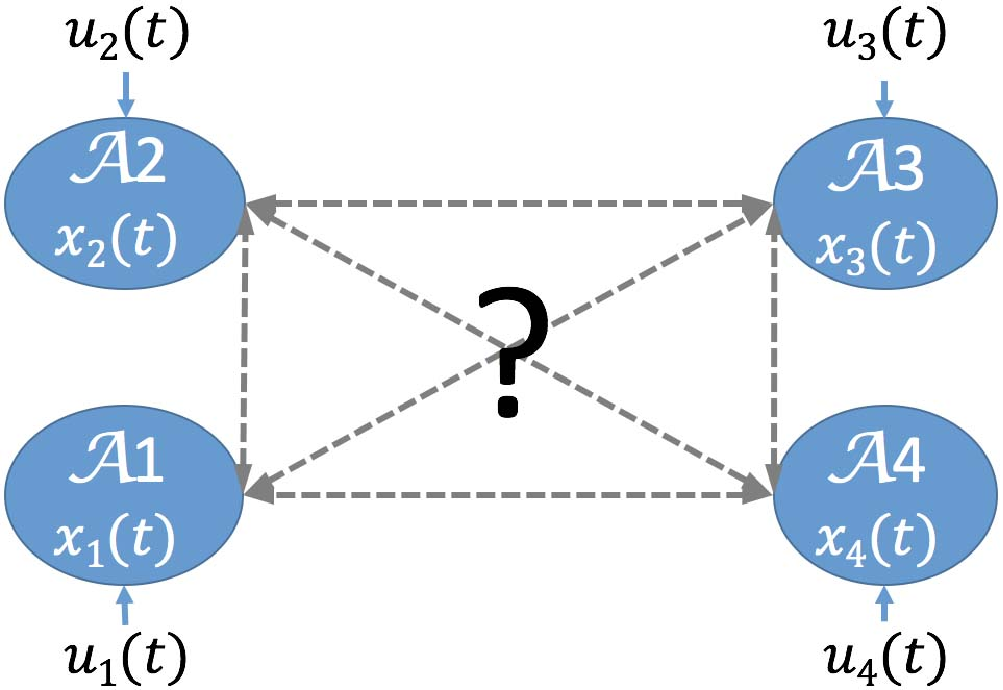}}\quad\quad\quad\quad
    \subfloat[Corresponding feedback signal]{\includegraphics[scale=0.35]{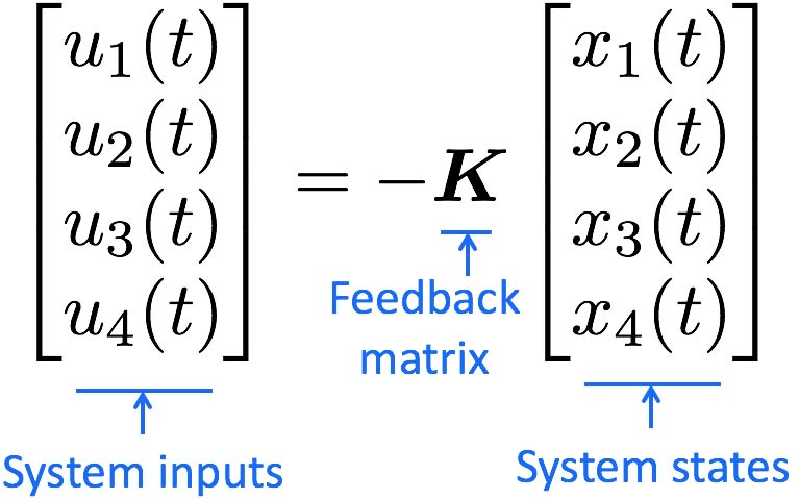}}
    \caption{Illustration of a distributed system having four agents denoted by $\cA1$, $\cA2$, $\cA3$ and $\cA4$, and its feedback signal model, where $\bx_i(t)$ and $\bu_i(t)$ are the state and input vectors of the $i$-th agent at time $t$ respectively and $\bK$ represents the feedback matrix of the system.}
    \label{fig:DS}
\end{figure}

Since the inputs and the states of the system are organized by stacking the inputs and the states of the multi-agents in the distributed system, non-zero elements in off-diagonal (or off-block-diagonal) of the feedback matrix $\bK$ is related to the communication link from one agent to another in the distributed system. For example, as shown in Fig. \ref{fig:DS} (a), suppose a distributed system has four agents, $\cA1$, $\cA2$, $\cA3$, and $\cA4$, where $\bx_i(t)$ and $\bu_i(t)$ are the state and input vectors of the $i$-th agent at time $t$ respectively. The state vector $\bx(t)$ and the input vector $\bu(t)$ for the whole distributed system are organized by stacking each state and input of the agent as shown in Fig. \ref{fig:DS} (b). If the $i$-th row and $j$-th column block of the feedback matrix $\bK$ has non-zero elements, then, it represents that the state of the $j$-th agent needs to be transmitted to the input of the $i$-th agent. Therefore, in order to reduce the communication links in the system, we consider sparsity constraints or regularization terms for sparse feedback matrix $\bK$.

For the given state space representation \eqref{eq:SSE_feedback}, the Linear Quadratic Regulator (LQR) objective function is defined as 
\par\noindent\small
\begin{align}\label{def_J0}
J(\bK)  & = \int_{t=0}^{\infty} \bx(t)^T \bQ \bx(t) + \bu(t)^T \bR \bu(t) \; dt \nonumber \\
			   & = \int_{t=0}^{\infty} \bx(t)^T ( \bQ + \bK^T \bR \bK) \bx(t) \; dt 
\end{align}
\normalsize
where $\bQ \succeq 0 \in \R^{n \times n}$ and $\bR \succ 0 \in \R^{m \times m}$ are given performance weight matrices. 
By introducing a matrix $\bP \in \R^{n \times n}$  such that $\frac{d}{dt} \bx(t)^T \bP \bx(t) = - \bx(t)^T (\bQ + \bK^T \bR \bK) \bx(t)$, $J(\bK)$ is further stated as
\par\noindent\small
\begin{align}
 J(\bK) & = \int_{t=0}^{\infty} -\frac{d}{dt} \bx(t)^T \bP \bx(t) dt = \bx(0)^T \bP \bx(0) = \Tr(\bP \bS_x),
\end{align} 
\normalsize
where $\lim_{t \rightarrow \infty} \bx(t) = 0$ due to the assumption that the feedback system is asymptotically stable, and we denote $\bx(0)\bx(0)^T$ as $\bS_x$. From the equation that $\frac{d}{dt} \bx(t)^T \bP \bx(t) = \dot{\bx}(t)^T \bP \bx(t) + \bx(t)^T \bP \dot{\bx}(t)$, where $\dot{\bx}(t) = (\bA - \bB_1\bK)\bx(t) + \bB_2 \bw(t)$ and $\E[\bw(t)] = \bm{0}$, for a matrix $\bK$, the matrix $\bP$ should satisfy the following Lyapunov equation:
\par\noindent\small
\begin{align} \label{eq:lya_P}
	& ( \bA - \bB \bK)^T \bP + \bP ( \bA - \bB \bK) + \bQ + \bK^T \bR \bK = \bm{0}.
\end{align}
\normalsize
From the definition of $\bP$ and $\bQ + \bK^T \bR \bK \succ 0$, $\bP$ is strictly positive definite. Additionally, for a given matrix $\bK \in \cF$, there exists a unique solution $\bP$. This is because all the eigenvalues of $\bA - \bB \bK$ are negative and the sum of any two eigenvalues of $\bA - \bB \bK$ is not zero, which indicates that $(\bI \otimes (\bA - \bB \bK)^T + (\bA -\bB \bK)^T \otimes \bI)$ is non-singular \cite{chen1998linear}. Note that \eqref{eq:lya_P} can be restated as $(\bI \otimes (\bA - \bB \bK)^T + (\bA -\bB \bK)^T \otimes \bI) \vect(\bP) = - \vect(\bQ +\bK^T \bR \bK)$,
where $\vect(\cdot)$ represents the vectorization operator for a matrix by stacking the columns of a matrix. 
Thus, $\bP$ can be thought of as a function of $\bK$, denoted by $\bP(\bK)$.

Then, the LQR minimization problem with a regularization term for $\bK$ is expressed as follows:
\par\noindent\small
\begin{align}\label{eq:LQR_prob}
	& \underset{\bK}{\text{minimize}} \; J(\bK) + \gamma G(\bK) \;\;\;\; \text{subject to} \;   \bK \in \cF,
\end{align}
\normalsize
where $J(\bK) =\Tr(\bP(\bK) \bS_x)$, $\bP(\bK)$ satisfies the Lyapunov equation \eqref{eq:lya_P},  and $G(\bK)$ is a regularization term for the structure of the feedback matrix $\bK$, e.g., $G(\bK) = \| \bK \|_1$ for a sparse feedback matrix in element-wise, and $\gamma \geq 0 $ is a weight parameter for the regularization term. 

In order to calculate the gradient of $J(\bK)$, we further introduce a Lyapunov equation having $\bS_x$ term as follows:
\par\noindent\small
\begin{align} \label{eq:lya_L} 
	( \bA - \bB \bK) \bL(\bK) + \bL(\bK) ( \bA - \bB \bK)^T + \bS_x = \bm{0},
\end{align} 
\normalsize
where a variable $\bL(\bK) \in \R^{n \times n}$, a function of $\bK$. We then have the gradient of $J(\bK)$ over $\bK$ from Theorem 3.2 of  \cite{rautert1997computational} as
\par\noindent\small
\begin{align}\label{eq:grad_J}
	\grad J(\bK)  = 2 [ \bB^T \bP(\bK) - \bR \bK ] \bL(\bK),
\end{align}
\normalsize
where $\bL(\bK)$ and $\bP(\bK)$ hold the Lyapunov equation \eqref{eq:lya_L} and \eqref{eq:lya_P} respectively. Additionally, for a matrix $\bK \in \cF$, $\bL(\bK)$ is also uniquely determined.

We can also consider $G(\bK)$ for the sparsity on the feedback matrix $\bK$ in \eqref{eq:LQR_prob} as a part of constraints as follows:
\par\noindent\small
\begin{align}\label{eq:LQR_prob_change}
	& \underset{\bK}{\text{minimize}} \; J(\bK) \nonumber \\
	& \text{subject to} \; \bK \in \cF, \;G(\bK)  \leq s_{\gamma}.
\end{align}
\normalsize
where $s_{\gamma} \geq 0$ is an upper bound of the sparsity-promoting function value. Our goal in this paper is finding sparse feedback controllers minimizing LQR by solving  the sparse optimal control design problems \eqref{eq:LQR_prob} and \eqref{eq:LQR_prob_change}.

In the next section, we will introduce the previous methods for sparse LQR control design including the method using the ADMM \cite{lin2013design} to solve \eqref{eq:LQR_prob} and the algorithm using the GraSP \cite{lian2017game} to solve \eqref{eq:LQR_prob_change}.

\section{Previous Research on Sparse LQR Control Design}
\label{sec:review}
The sparse optimal control design problem introduced in \eqref{eq:LQR_prob} has been studied in \cite{lin2013design,fardad2009optimal}. Especially, the authors in \cite{lin2013design} considered the ADMM by separately minimizing $J(\bK)$ and $G(\bK)$ in the ADMM iterative steps. For the ADMM update, \eqref{eq:LQR_prob} is restated as follows by introducing a new variable $\bF \in \R^{m \times n}$ for variable separation:
\par\noindent\small
\begin{align}\label{eq:LQR_ADMM}
	& \underset{\bK,\;\bF}{\text{minimize}} \; J(\bK) + \gamma G(\bF) \;\;\text{subject to} \; \bK \in \cF, \;\bK = \bF.
\end{align}
\normalsize
Note that due to the last equality constraint, i.e., $\bK = \bF$, \eqref{eq:LQR_ADMM} is equivalent to \eqref{eq:LQR_prob}. Then, from \eqref{eq:LQR_ADMM}, we can have the augmented Lagrangian as follows:
\par\noindent\small
\begin{align*}
	& L_{\rho}(\bK,\bF,\bLambda) := J(\bK) + \gamma G(\bF) + \langle \bK - \bF, \bLambda \rangle + \frac{\rho}{2} \| \bK - \bF \|_F^2,
\end{align*}
\normalsize
where $\rho \geq 0$  is a parameter value for the augmented term and $\bLambda$ is a dual variable. A small $\rho$ value allows the ADMM to search possible solutions far-away from the current point. 

With this augmented Lagrangian, updating variables $\bK$, $\bF$, and $\bLambda$ is conducted as follows:
\par\noindent\small
\begin{align}
	&\bK^{(t+1)} = \argmin_{\bK \in \cF} \; L_{\rho} (\bK,\bF^{(t)},\bLambda^{(t)}),\label{eq:ADMM_step1}\\
	&\bF^{(t+1)} = \argmin_{\bF} \; L_{\rho} (\bK^{(t+1)},\bF,\bLambda^{(t)}),\label{eq:ADMM_step2}\\
	&\bLambda^{(t+1)} = \bLambda^{(t)} + \rho(\bK^{(t+1)} - \bF^{(t+1)}) \label{eq:ADMM_step},	
\end{align}
\normalsize
where the super-script $(t)$ represents the $t$-th iteration. The two Lyapunov equations for $\bP(\bK)$ and $\bL(\bK)$ and the feasible set constraint $\bK \in \cF$ are considered in updating $\bK^{(t+1)}$ by employing Anderson-Moore algorithm \cite{rautert1997computational}. For the sparse-promoting function $G(\bF)$, various functions including $\ell_1$ norm, block $\ell_1$ norm, and sum of log functions, plus their weighted functions were addressed in \cite{lin2013design}.

Unlike the previous work \cite{lin2013design} to solve \eqref{eq:LQR_prob}, the authors in \cite{lian2017game} considered the sparsity condition as a part of constraints of the LQR minimization problem as in \eqref{eq:LQR_prob_change}. Especially, the $\ell_0$ norm of the off-diagonal of the feedback matrix is considered in \cite{lian2017game} for the sparsity-promoting function; namely, $G(\bK) = \sum_{i\neq j} \| \bK_{i,j} \|_0$, where $\bK_{i,j}$ is the $i$-row and the $j$-th column block of the matrix $\bK$ and $\| \bK_{i,j} \|_0$ is the number of non-zero elements in the partial feedback matrix $\bK_{i,j}$. In order to solve \eqref{eq:LQR_prob_change} with $\ell_0$ norm of the off-diagonal of the matrix $\bK$, the Gradient Support Pursuit (GraSP) \cite{bahmani2013greedy} is taken into account in \cite{lian2017game}. In the algorithm, the gradient descent is conducted as follows:
\par\noindent\small
\begin{align*}
 \bK^{(t+1)} = \bK^{(t)} - \rho^{(t)} \bigtriangleup J(\bK^{(t)}),
 \end{align*}
 \normalsize
where $\bigtriangleup J(\bK^{(t)})$ represents the descent direction of the objective function $J(\bK)$. \textcolor[rgb]{0,0,0}{In \cite{lian2017game}, the authors used Newton method to have the descent direction.} And then, a sparse feedback matrix is obtained by only maintaining the $s$ off-diagonal elements having the largest absolute values in $\bK^{(t+1)}$ and making other off-diagonal elements to zero. The non-zero support of the sparsity-promoted matrix $\bK$, which is the index set indicating the locations of non-zero elements in $\bK$, is updated by merging the previous support and the current support. Thanks to the truncating operation, which can be understood as the projection onto the set $\sum_{i \neq j} \| \bK_{i,j} \|_0 \leq s_{\gamma}$, the GraSP based algorithm successfully provides a sparse feedback matrix $\bK$. 

In the next sections, we will introduce the simple but efficient algorithms that we propose to solve the sparse-optimal control design problems \eqref{eq:LQR_prob} and \eqref{eq:LQR_prob_change}.

\section{Iterative Shrinkage-Thresholding Algorithm}
\label{sec:algo}
In the field of signal processing, for sparse signal recovery, the ISTA was well studied in the previous research \cite{beck2009fast,daubechies2004iterative,tibshirani2013lasso} and references therein. Inspired by the conventional ISTA in signal recovery, we consider to use the algorithm concept to the sparse LQR optimal control design problem. However, due to the constraints in \eqref{eq:LQR_prob} and \eqref{eq:LQR_prob_change}, in the context of optimization, the LQR optimal control problem is understood as a non-convex optimization problem due to feasible set $\cF$ and bilinear Lyapunov equation, while the sparse signal recovery problem considered in \cite{beck2009fast,daubechies2004iterative,tibshirani2013lasso} is a convex optimization problem. Therefore, in this section, we introduce how to tackle the sparse LQR optimal control design problem, which is a non-convex optimization problem, by using the concept of ISTA. Before introducing that, let us review the conventional ISTA for sparse signal recovery.

\subsection{Description of conventional iterative shrinkage-thresholding algorithm for sparse signal recovery}
\label{subsec:conv_ISTA}
The linear inverse problem in signal processing is the problem estimating an unknown signal $\bx \in \R^{n}$ from the linear measurement $\by \in \R^{m}$ with a sensing matrix $\bH \in \R^{m \times n}$; namely, $\by = \bH \bx + \be$, where $\be \in \R^m$ is a noise vector. For this problem, the Least Square (LS) minimization was studied to estimate the unknown signal $\bx$ with minimum of data error in $\ell_2$ norm as follows:
\par\noindent\small
\begin{align}
		& \underset{\bx \in \R^{n} }{\text{minimize}} \; || \by - \bH \bx ||^2_2.
\end{align}
\normalsize
If $m \geq n$ and $\bH^T \bH$ is non-singular, then the solution of the LS minimization problem is obtained as $\bx^{\star} = (\bH^T \bH)^{-1} \bH^T \by$. However, if $m < n$, then, the LS minimization can possibly have many solutions and become ill-conditioned.

By considering additional conditions on the unknown signal $\bx$ such as sparsity or block sparsity, we can have a unique solution through the following Least Absolute Shrinkage and Selection Operator (LASSO) optimization problem \cite{beck2009fast,daubechies2004iterative,tibshirani2013lasso}:
\par\noindent\small
\begin{align}\label{eq:lasso}
		& \underset{\bx \in \R^{n} }{\text{minimize}} \; \frac{1}{2}\| \by - \bH \bx \|^2_2 + \gamma \| \bx \|_1,
\end{align}
\normalsize
where $\by \in \R^{m}$ is a measurement vector, $\bH \in \R^{m \times n}$, $m< n$, is a sensing matrix and  $\gamma \geq 0$ is a tuning parameter for the sparsity level of $\bx$. Here, $\ell_1$ norm is considered for the sparsity-promotion on the unknown vector $\bx$, which is the sum of absolute value of each element of  $\bx$, i.e., $\|\bx \|_1 = \sum_{i=1}^n | x_i |$. 

In order to solve the LASSO problem \eqref{eq:lasso}, the researchers including in \cite{beck2009fast,daubechies2004iterative} proposed the Iterative Shrinkage-Thresholding Algorithm (ISTA) for sparse signal recovery, where the updating step of the ISTA is stated as follows:
\par\noindent\small
\begin{align}\label{eq:ISTA_conv}
	\bx^{(t+1)}  = \cS_{\gamma \mu } \bigg( \bx^{(t)} + 2\mu \bH^T ( \by - \bH \bx^{(t)}) \bigg),
\end{align}
\normalsize
where the super-script $(t)$ represents the $t$-th iteration, and $\mu$ is the step size of the gradient descent.  The shrinkage-threshold operator $\cS_a(\cdot): \R^{n} \rightarrow \R^{n}$ is stated as follows:
\par\noindent\small
\begin{align} \label{eq:threshold_operator}
	\cS_{a}(\bx)_i = \sgn(x_i) \max \{ |x_i| - a , \; 0\},
\end{align}
\normalsize
where $\sgn(\cdot)$ is the sign function and $a \geq 0$. The convergence rate of the ISTA in terms of the objective value gap between the optimal objective value and the objective value at the $t$-th iteration is known to be $O(\frac{1}{t})$, where $t$ is the number of iterations. 

In \cite{beck2009fast}, the Fast ISTA (FISTA) was introduced with faster convergence speed. By using Nesterov's accelerated gradient descent \cite{nesterov2003introductory}, which is a techniqe using momentum of the gradient descent, the FISTA achieves $O(\frac{1}{t^2})$ convergence rate in terms of the objective value gap. The updating steps of the FISTA are stated as follows:
\par\noindent\small
\begin{align}
	& 	\bx^{(t+1)}  = \cS_{\gamma \mu } \bigg( \tilde{\bx}^{(t)} + 2\mu \bH^T ( \by - \bH \tilde{\bx}^{(t)}) \bigg), \nonumber \\
	& \tilde{\bx}^{(t+1)} = (1- \beta^{(t)}) \bx^{(t+1)} + \beta^{(t)} \bx^{(t)} \label{eq:FISTA_step},	
\end{align}
\normalsize
where $\alpha^{(0)} = 0$, $\alpha^{(t)} = \frac{1+ \sqrt{1+4 {\alpha^{(t-1)}}^2} }{2}$, and $\beta^{(t)} = \frac{1- \alpha^{(t)}}{\alpha^{(t+1)}}$.

Then, we will introduce our proposed method on solving the sparse LQR optimal control design problem by using the concept of ISTA.

\subsection{Iterative shrinkage-thresholding algorithm for sparse LQR optimal control design} 
\label{subsec:algo2}
The major difficulty in applying the conventional ISTA to the sparse optimal control design problem is that we have the constraint for $\bK$ to be in the feasible set $\cF$ as in \eqref{eq:LQR_prob} and the Lyapounv equations \eqref{eq:lya_P} and \eqref{eq:lya_L}, while the LASSO problem \eqref{eq:lasso} has no constraint. Hence, with the simple updating scheme \eqref{eq:ISTA_conv}, the LASSO problem can be solved.  However, in the sparse optimal control design problem, which is a non-convex optimization problem, due to the feasible conditions including $\cF$, solving \eqref{eq:LQR_prob} by using the concept of the thresholding algorithms is not trivial.

To derive updating step, for a current point $\bK^{(t)}$, we can upper-bound $J(\bK)$, where $\bK \in \cF$, with appropriately chosen $\rho^{(t+1)}$ as follows:
\par\noindent\small
\begin{align}\label{eq:UB_JK}
	J(\bK) & \hspace{-0.1em}\leq \hspace{-0.1em}J(\bK^{(t)}) \hspace{-0.1em}+\hspace{-0.1em} \langle \bK \hspace{-0.1em}- \hspace{-0.1em}\bK^{(t)} , \grad J(\bK^{(t)}) \rangle \hspace{-0.1em}+\hspace{-0.1em} \frac{\rho^{(t+1)}}{2} \| \bK \hspace{-0.1em}- \hspace{-0.1em}\bK^{(t)}\|_F^2 \nonumber \\
	&\hspace{-0.1em} := \hspace{-0.1em}\tilde{J}(\bK,\bK^{(t)},\rho^{(t+1)}), 
\end{align}
\normalsize
where $\rho^{(t+1)} \geq 0$ is a parameter to achieve the bound. Then, we can have the iteration step by minimizing the upper-bound of $J(\bK) + \gamma G(\bK)$ as follows:
\par\noindent\small
\begin{align}
	\bK^{(t+1)} & \hspace{-0.2em}= \hspace{-0.1em}\argmin_{\bK\in \cF} \; \hspace{-0.1em}\tilde{J}(\bK,\bK^{(t)},\rho^{(t+1)}) + \gamma G(\bK) \label{eq:update_K}\\
					 & \hspace{-0.2em}= \hspace{-0.1em}\argmin_{\bK\in \cF} \; \hspace{-0.1em}\gamma G(\bK) \hspace{-0.1em} + \hspace{-0.1em} \frac{\rho^{(t+1)}}{2} \| \bK \hspace{-0.2em}- \hspace{-0.2em}\bK^{(t)} \hspace{-0.2em}+ \hspace{-0.1em}\frac{1}{\rho^{(t+1)}} \grad \hspace{-0.1em}J(\bK^{(t)}) \|_F^2.  \nonumber
\end{align}
\normalsize

Depending on the sparsity-promoting regularization term $G(\bK)$, we can have different updating step for $\bK^{(t+1)}$. For element-wise sparse solution of $\bK \in \R^{m \times n}$, we set
$	G(\bK) = \| \bK \|_1 = \sum_{l=1}^{m} \sum_{k=1}^{n} |K_{l,k}|$, where $K_{l,k}$ represents the $l$-th row and $k$-th column element of $\bK$. Then, we have the following updating step for $\bK^{(t+1)}$:
\par\noindent\small
\begin{align}\label{eq:update_l1}
	\bK^{(t+1)}  & = \cS_{\frac{\gamma}{\rho^{(t+1)}}} \bigg(\bK^{(t)} - \frac{1}{\rho^{(t+1)}} \grad J(\bK^{(t)} ) \bigg)
\end{align}
\normalsize
where $\rho^{(t+1)}$ is chosen for satisfying $\bK^{(t+1)} \in \cF$ and  \eqref{eq:UB_JK} via backtracking operation and $\cS_a(\bb)$ represents the shrinkage-threshold operator introduced in \eqref{eq:threshold_operator}. \textcolor[rgb]{0,0,0}{For readability, let us introduce the backtracking operation below \eqref{eq:update_blkw1}.} 

If we consider the block sparsity on off-diagonal of $\bK$, we can use the block $\ell_1$ norm for the regularization term $G(\bK)$ as
	$G(\bK) = \sum_{i,j: i \neq j} \| \bK_{i,j} \|_F$,
where $\bK_{i,j}$ represents the $i$-th row and the $j$-th column block of $\bK$. Then, we have the following updating step for $\bK^{(t+1)}$:
\par\noindent\small
\begin{align*}
	\bK^{(t+1)}  
					 & =  \argmin_{\bK \in \cF} \; \sum_{i,j: i\neq j} \bigg( \gamma \| \bK_{i,j} \|_F  \\
					&\quad\;\;+  \frac{\rho^{(t+1)}}{2} \| \bK_{i,j} - \bK^{(t)}_{i,j} + \frac{1}{\rho^{(t+1)}} \grad J(\bK^{(t)})_{i,j} \|_F^2 \bigg).
\end{align*}
\normalsize
For the $i$-th row and the $j$-th column block, where $i \neq j$, we have 
\par\noindent\small
\begin{align}\label{eq:update_blkl1}
	\bK^{(t+1)}_{i,j} 
							& =  \cB_{ \frac{ \gamma}{ \rho^{(t+1)} } } \bigg( \bK^{(t)}_{i,j} - \frac{1}{\rho^{(t+1)}} \grad J(\bK^{(t)})_{i,j} \bigg),
\end{align}
\normalsize
where $\cB_{a}(\bb)$ is the block shrinkage-threshold operator defined as follows: 
\par\noindent\small
\begin{align}\label{eq:blk_thre}
	\cB_{a}(\bb) = \max\bigg\{\| \bb \|_F - a, 0 \bigg\}\cdot\frac{\bb}{\| \bb \|_F}.
\end{align}
\normalsize

For better element-wise sparsity on $\bK$, we can also consider an iterative reweighted algorithm with $G(\bK)$ defined as
$	G(\bK) = \sum_{l=1}^{m} \sum_{k=1}^{n} W_{l,k} | K_{l,k} |$, where $W_{l,k}$ is the weight value for the the $l$-th row and the $k$-th column element of the matrix $\bK$. The weight for computing $K^{(t+1)}_{l,k}$, denoted by $W^{(t)}_{l,k}$, is defined as follows:
	$W^{(t)}_{l,k} = \frac{1}{ | K_{l,k}^{(t)} | + \epsilon}$. Here $\epsilon$ is a small positive constant to prevent the weight value to be infinity. For the $l$-th row and $j$-th column element of the feedback matrix $\bK$ at the $(t+1)$-th iteration, we  have
\par\noindent\small
\begin{align}\label{eq:update_wl1}
	K^{(t+1)}_{l,k} = \cS_{\frac{\gamma W^{(t)}_{l,k}}{\rho^{(t+1)}}} \bigg( K^{(t)}_{l,k} - \frac{1}{\rho^{(t+1)}} \grad J(\bK^{(t)} )_{l,k} \bigg),
\end{align}
\normalsize
where $l=1,2,...,m$ and $k=1,2,...,n$.

For better block sparsity operator, by having $G(\bK) = \sum_{i,j} W_{i,j}\| \bK_{i,j} \|_F$, where $W_{i,j}$ is the weight value for the the $i$-th row and the $j$-th column block of $\bK$, calculated as $W^{(t)}_{i,j} = \frac{1}{\| \bK_{i,j}^{(t)} \|_F + \epsilon}$. Then, for the $i$-th row and $j$-th column block of $\bK^{(t+1)}$, we have 
\par\noindent\small
\begin{align}\label{eq:update_blkw1}
	\bK^{(t+1)}_{i,j} =  \cB_{ \frac{ \gamma W_{i,j}^{(t)}}{ \rho^{(t+1)} } } \bigg( \bK^{(t)}_{i,j} - \frac{1}{\rho^{(t+1)}} \grad J(\bK^{(t)})_{i,j} \bigg).
\end{align}
\normalsize


\textcolor[rgb]{0,0,0}{It is noteworthy that the inside of the thresholding operators $\cS(\cdot)$ and $\cB(\cdot)$ is the conventional gradient descent operation at the $t$-th iteration with the step size $\frac{1}{\rho^{(t+1)}}$.  Therefore, the operations \eqref{eq:update_l1}, \eqref{eq:update_blkl1}, \eqref{eq:update_wl1}, and \eqref{eq:update_blkw1} can be understood as the combined operations - gradient descent operation and threshold operation for sparsity. Since the $(t+1)$-th iteration solution $\bK^{(t+1)}$ also needs to be in $\cF$ for the stability of the close-loop system, after obtaining the $(t+1)$-th iteration result, we check whether $\bK^{(t+1)}$ is in $\cF$ or not by investigating the maximum of the real part of eigenvalue of the matrix $\bA - \bB\bK^{(t+1)}$. If $\bK^{(t+1)}$ is in $\cF$, then, we conduct the same updating step for the next iteration value. If $\bK^{(t+1)}$ is not in $\cF$, then, we reduce the step size $\frac{1}{\rho^{(t+1)}}$ by increasing $\rho^{(t+1)}$ and search again the optimal feedback matrix $\bK^{(t+1)}$ along the descent direction $\grad J(\bK)$ at $\bK^{(t)}$. Remark that if $\bK^{(t)}$ is in $\cF$, then, there is $\rho^{(t+1)}$ satisfying $\bK^{(t+1)} \in \cF$ and  \eqref{eq:UB_JK}. This is because as $\rho^{(t+1)}$ goes to infinity, the step size $\frac{1}{\rho^{(t+1)}}$ as well as the threshold level of the sparsity operator, i.e., $\frac{\gamma}{\rho^{(t+1)}}$ go to 0. In the worst case when $\rho^{(t+1)} = \infty$, we have the same result $\bK^{(t)} \in \cF$ for the $(t+1)$-th iteration result, i.e., $\bK^{(t+1)} = \bK^{(t)} \in \cF$. Thus, we can find a solution in $\cF$ with a backtracking operation.} 
Finally, after obtaining $\bK^{(t+1)}$, $\bK^{(t)} \in \cF$, if the Euclidean distance between $\bK^{(t+1)}$ and $\bK^{(t)}$ is less than a certain predetermined small tolerant value $\epsilon$, we stop the algorithm. If the algorithm is not converged within the predetermined maximum iterations, denoted as $MaxItr$, then, we stop the algorithm and provide the final result for a structured feedback matrix $\bK$. Algorithm \ref{alg:ISTA} describes the whole detail steps to solve the sparse LQR optimal control design problem \eqref{eq:LQR_prob}. 

\begin{algorithm}[t]
  \caption{Iterative Shrinkage Thresholding Algorithm (ISTA) for Sparse LQR Optimal Control Design}
  \label{alg:ISTA}
  \SetAlgoLined
  \SetKwRepeat{Do}{do}{while}%
{\small
   \KwIn{ $\bA$, $\bB$, $\bS_x$, $\bQ \succeq 0$, $\bR \succ 0$, $\rho_{0}$, $\gamma$, $\alpha > 1$, $\epsilon$, $MaxItr$  }
   \textbf{Initialization}: $\bK^{(0)} \leftarrow$ a stable dense matrix $\bK \in \cF$ from a solution to LQR problem with $\gamma = 0$, $\rho^{(1)} \leftarrow \rho_{0}$   \par
   \For { $t=0$  \KwTo $MaxItr$ }
   {
   		   $\bP^{(t)}$ $\leftarrow$ solution of \eqref{eq:lya_P} from $\bK^{(t)}$ \par
   		   $\bL^{(t)}$ $\leftarrow$ solution of \eqref{eq:lya_L} from $\bK^{(t)}$ \par
   		   $\grad J(\bK^{(t)}) $ $\leftarrow$ calculating the gradient value of $J(\bK)$ at $\bK^{(t)}$ from \eqref{eq:grad_J}  \par
   		   \Do{ $\rho^{(t+1)} \neq \rho_{0}$ }{
	             $\bK^{(t+1)}$ $\leftarrow$ solution of \eqref{eq:update_K} \par
		   		\uIf { $\max(\Re(\lambda(\bA - \bB\bK^{(t+1)})) \geq 0$  \textcolor[rgb]{0,0,0}{or} $J(\bK^{(t+1)}) > \tilde{J}(\bK^{(t+1)},\bK^{(t)}, \rho^{(t+1)} )$}    
		   		{
		   			$\rho^{(t+1)}$ $\leftarrow$ $\rho^{(t+1)} \times \alpha$ \Comment{Backtracking with a smaller step-size} \par
		   		}\Else {
					$\rho^{(t+1)}$ $\leftarrow$ $\rho_{0}$ \par		
				}
			}
			\If { $\| \bK^{(t+1)} - \bK^{(t)} \|_F < \epsilon$ } 
			{
				break
			} 			
}
  \KwOut{ Structured feedback matrix $\bK$ }
  }
\end{algorithm}


\subsection{Comparisons and merits}
\label{subsec:merits}
The ISTA for the sparse LQR optimal control design is different from the conventional ISTA introduced in the field of signal processing in two aspects - feasible set and Lyapunov equation. The feasible set of the conventional ISTA is normally $\R^n$, where $n$ is the signal dimension, while we have the feasible set $\cF$ in order for the close-loop system to be asymptotically stable, which is known to be non-convex. Additionally, at each iteration, the Lyapunov equations are necessary to be solved, which is a bilinear function,  in order to calculate the gradient at $\bK^{(t)}$, while the conventional ISTA does not have those additional equations to solve. Therefore, in the context of optimization, the sparse LQR optimal control problems introduced in \eqref{eq:LQR_prob} and \eqref{eq:LQR_prob_change} are understood as constrained non-convex optimization problems, while the LASSO problem in \eqref{eq:lasso} is a unconstrained convex optimization problem.

In addition, the benefits of our algorithm against the ADMM-based algorithm introduced in \cite{lin2013design} are two-fold. Firstly, our algorithm is simpler than the ADMM-based one. More specifically, the updating variables at each iteration in our algorithm are $\bK$, $\bP$, and $\bL$, while ADMM includes additional primal variable and dual variable. Hence, it is more complex than our algorithm, and requires more memory space at each iteration. Also, thanks to the simplicity of our algorithm, it can be easily adapted to its model-based DNN implementation, which we will introduce in Section \ref{sec:DNN}. It is noteworthy that it may also be possible to implement a model-based DNN based on the ADMM with additional network complexity and memory usage. The second advantage of our proposed algorithm against ADMM \cite{lin2013design} is that our proposed algorithm can be faster than the ADMM in the execution time for the same sparsity level of the feedback matrix $\bK$ and LQR cost. Throughout various numerical experiments, we will demonstrate the better performance of our proposed algorithm in the numerical experiment section \ref{sec:experiment}.

\textbf{Remark:} In terms of operations in comparison, the updating steps \eqref{eq:ADMM_step1} and \eqref{eq:ADMM_step2} in ADMM can be seen as the calculation of \eqref{eq:update_l1} (or \eqref{eq:update_blkl1}, \eqref{eq:update_wl1}, or \eqref{eq:update_blkw1}) in ISTA. The fixed-point operation, i.e., Anderson-Moore algorithm \cite{rautert1997computational}, in ADMM can  be compared to the backtracking operation in ISTA. The comparison of complexity between ISTA and ADMM is not straight-forward due to the presence of an inner loop for the fixed point operation in ADMM and for the backtracking operation in ISTA. From a big-O notation perspective, we believe that there is no significant difference in complexity between the ISTA and the ADMM-based approach. However, when designing model-based deep neural networks using ISTA, we can highlight clear advantages over ADMM.

\subsection{Analysis of Iterative Shrinkage-Thresholding Algorithm for Sparse LQR Optimal Control Design}
\label{sec:analysis}

Even though the convergence analysis of the conventional ISTA is well-known, the convergence analysis of the ISTA for the sparse optimal control design is unclear due to the aforementioned differences. In this sub-section, we introduce the analyses of the ISTA for the sparse LQR optimal control design problem. 

We start our algorithm from a starting point $\bK^{(0)} \in \cF$, which is the solution to the conventional LQR problem, i.e.,  \eqref{eq:LQR_prob} with $\gamma =0$. It is well-known in the control field that $\bK^{(0)}$ can be obtained by solving the following Riccati equation over $\bP$ \cite{rautert1997computational}:
\par\noindent\small
\begin{align}
	\bA^T \bP + \bP \bA - \bP \bB \bR^{-1} \bB^T \bP + \bQ = 0.
\end{align}
\normalsize
Then,  $\bK^{(0)} = \bR^{-1}\bB^T \bP$. 

Let us denote $F(\bK) := J(\bK) + \gamma G(\bK)$ and the upper bound of $J(\bK) + \gamma G(\bK)$ introduced in \eqref{eq:update_K}  by $\tilde{F}(\bK,\bK^{(t)},\rho)$; namely,
\par\noindent\small
\begin{align}\label{def:F_tilde}
&\tilde{F}(\bK,\bK^{(t)}, \rho)   \\
& := \hspace{-0.1em}  J(\bK^{(t)})\hspace{-0.1em}  + \hspace{-0.1em} \langle   \bK \hspace{-0.1em} - \hspace{-0.1em}  \bK^{(t)}\hspace{-0.1em}  , \hspace{-0.1em} \grad J(\bK^{(t)})  \rangle \hspace{-0.1em} + \hspace{-0.1em}  \frac{\rho}{2} \| \bK\hspace{-0.1em}  - \hspace{-0.1em} \bK^{(t)} \|_F^2 \hspace{-0.1em} +\hspace{-0.1em}  \gamma G(\bK).\nonumber
\end{align}
\normalsize
Remark that at the $t$-th iteration, the ISTA minimizes $\tilde{F}(\bK,\bK^{(t)}, \rho^{(t+1)})$. We then have the following lemma on the ISTA for the sparse optimal control design problem \eqref{eq:LQR_prob}. 

\begin{lemma}\label{thm:decreasing_func}
Given $\rho_{0}$ and any initial point $\bK^{(0)} \in \cF$, Algorithm \ref{alg:ISTA} provides the non-increasing objective value, namely:
\par\noindent\small
\begin{align}
	F(\bK^{(t+1)}) \leq  F(\bK^{(t)})
\end{align}
\normalsize
\end{lemma}
\begin{IEEEproof}
In the proof, we will show 
\par\noindent\small
\begin{align*}
 F(\bK^{(t+1)}) & = \tilde{F}(\bK^{(t+1)}, \bK^{(t+1)}, \rho^{(t+1)}) \\
 					  & \leq \tilde{F}(\bK^{(t+1)}, \bK^{(t)}, \rho^{(t+1)})  \\
 					  & \leq \tilde{F}(\bK^{(t)}, \bK^{(t)}, \rho^{(t+1)}) = F(\bK^{(t)}),
\end{align*}
\normalsize
where the first equality is obtained from that, for any $\rho$, 
\par\noindent\small
\begin{align}
	\tilde{F}(\bK^{(t)}, \bK^{(t)}, \rho) & = J(\bK^{(t)}) + \langle   \bK^{(t)} - \bK^{(t)} , \grad J(\bK^{(t)})  \rangle \nonumber  \\
													   & \quad\quad +  \frac{\rho}{2} \| \bK^{(t)} - \bK^{(t)} \|_F^2 + \gamma G(\bK^{(t)}) \label{eq:F_tilda_at_Kt} \\ 
	& = J(\bK^{(t)}) + \gamma G(\bK^{(t)})  \nonumber \\
	& = F(\bK^{(t)}) \nonumber.
\end{align}
\normalsize
Then, the second inequality, $\tilde{F}(\bK^{(t+1)}, \bK^{(t)}, \rho^{(t+1)})  \leq \tilde{F}(\bK^{(t)}, \bK^{(t)}, \rho)$, is obtained with the following reason. Since \eqref{eq:F_tilda_at_Kt} holds for any $\rho$, $\tilde{F}(\bK^{(t)}, \bK^{(t)}, \rho^{(t+1)}) = F(\bK^{(t)})$. With the properly chosen step-size $1/\rho^{(t+1)}$ through the backtracking operation in Algorithm \ref{alg:ISTA}, we obtain $\bK^{(t+1)} \in \cF$, by minimizing \eqref{def:F_tilde}. Thus, we can have $\tilde{F}(\bK^{(t+1)}, \bK^{(t)}, \rho^{(t+1)})  \leq \tilde{F}(\bK^{(t)}, \bK^{(t)}, \rho^{(t+1)})$. Finally, the first inequality is obtained by showing $\tilde{F}(\bK^{(t+1)}, \bK^{(t)}, \rho^{(t+1)}) - \tilde{F}(\bK^{(t+1)}, \bK^{(t+1)}, \rho^{(t+1)}) \geq 0$ as
\par\noindent\small
\begin{align*}
	& \tilde{F}(\bK^{(t+1)}, \bK^{(t)}, \rho^{(t+1)}) - \tilde{F}(\bK^{(t+1)}, \bK^{(t+1)}, \rho^{(t+1)})  \\
	& = J(\bK^{(t)}) + \langle   \bK^{(t+1)} - \bK^{(t)} , \grad J(\bK^{(t)})  \rangle \\
	& \quad\quad +  \frac{\rho^{(t+1)}}{2} \| \bK^{(t+1)} - \bK^{(t)} \|_F^2  - J (\bK^{(t+1)})  \geq 0,
\end{align*}
\normalsize
where  the last inequality is obtained from the condition to check during the iteration in Algorithm \ref{alg:ISTA}.
\end{IEEEproof}

\begin{corollary}\label{thm:level_set}
Let $\cF$ be non-empty and denote the level set at $\bK^{(t-1)}$, where $t \geq 1$, as follows:
\par\noindent\small
\begin{align}
	\cR(\bK^{(t-1)}) := \{  \bK \in \cF \; |\; F(\bK) \leq F(\bK^{(t-1)})\}.
\end{align}
\normalsize
Then, given any matrix $\bK^{(0)} \in \cF$, for the sparse LQR optimal control design problem \eqref{eq:LQR_prob}, Algorithm \ref{alg:ISTA} always provides $\bK^{(t)} \in \cR(\bK^{(t-1)})$, where $\bK^{(t)}$ is the $t$-th iteration result.
\end{corollary}
\begin{IEEEproof}
At the $t$-th iteration, where $t \geq 1$, if $\bK^{(t)}$ is not in $\cF$, the algorithm keeps decreasing the step-size $1/\rho^{(t)}$ and threshold-level $\gamma/\rho^{(t)}$, by choosing the larger $\rho^{(t)}$, and conducting the gradient descent and threshold operation at the current point $\bK^{(t-1)}$ until finding $\bK^{(t)}$ in $\cF$ through backtracking operation. In the worst case, if $\rho^{(t)}$ becomes $\infty$, then $\bK^{(t)}$ becomes the same as $\bK^{(t-1)}$, which is in $\cR(\bK^{(t-1)})$. Therefore, with Lemma \ref{thm:decreasing_func}, the $t$-th iteration solution of Algorithm \ref{alg:ISTA} is always $\bK^{(t)} \in \cR(\bK^{(t-1)})$.
\end{IEEEproof}

From Lemma \ref{thm:decreasing_func}, we know that due to the gradient descent operation with modified step-size, the objective function value non-increases as the iteration of the algorithm goes. However, its convergence rate is not considered. Theorem \ref{thm:convergence} describes the convergence rate for Algorithm \ref{alg:ISTA}. 
\begin{theorem}\label{thm:convergence}
Given $\rho_{0}$ and any initial point $\bK^{(0)} \in \cF$, let $\bK^{\star}$ be a local optimal solution to \eqref{eq:LQR_prob} and $\{ \bK^{(t)} \}_{t=1}^{\infty}$ be the estimated results obtained from Algorithm \ref{alg:ISTA}. If $J: \R^{m \times n} \rightarrow \R$ is $\rho_{max}$-smooth, we then have the following error bound: 
\par\noindent\small
\begin{align}\label{eq:convergence}
	F(\bK^{(t)}) - F(\bK^{\star}) \leq \frac{\rho_{max} \| \bK^{(0)} - \bK^{\star} \|_F^2 }{2t },
\end{align}
\normalsize
where $\rho_{max}$ is the maximum value among $\rho^{(t)}$'s, $\forall t$.
\end{theorem}
\begin{IEEEproof}
Even though we have the feasible set $\cF$, the intermediate result $\bK^{(t)}$ is found in $\cF$ via backtracking operation, which is the same case of ISTA with backtracking introduced in \cite[Theorem 3.1]{beck2009fast}, where Lipschitz constant is unknown. Based on \cite[Theorem 3.1]{beck2009fast}, the convergence rate of ISTA for the sparse optimal control design problem is the same as \eqref{eq:convergence}.
\end{IEEEproof}


Unlike the sparse LQR optimal control design problem \eqref{eq:LQR_prob} which the ISTA and the ADMM tackle, we can consider the sparsity-promoting function $G(\bK)$ as a part of constraints of the optimization problem as in \eqref{eq:LQR_prob_change}. In the next section, we will introduce the algorithm that solves the sparse LQR optimal control design problem \eqref{eq:LQR_prob_change}.

\section{Iterative Sparse Projection Algorithm for Sparse LQR Optimal Control Design} 
\label{sec:ISPA}
The authors in \cite{lian2017game} addressed the optimization problem  \eqref{eq:LQR_prob_change} especially with $\ell_0$ norm of the feedback matrix $\bK$, and proposed the GraSP with $\ell_0$ norm sparsity-promoting function, which can be interpreted as the projection on the $\ell_0$-ball. In this section, we introduce the Iterative Sparse Projection Algorithm (ISPA) for the sparse LQR optimal control design problem, which can be understood as a generalized algorithm of the GraSP \cite{lian2017game}. Before introducing the ISPA, let us introduce the project onto the sparsity-promoting ball.

\subsection{Projection onto the sparsity-promoting ball} \label{subsec:proj}
Using the projection operation onto the sparsity-promoting ball, e.g., $\ell_1$ ball, in an algorithm to obtain a sparse vector or matrix has been studied in the various fields such as machine learning \cite{duchi2008efficient} and signal processing \cite{starck2010sparse}. With the constraint $G(\bK) \leq s$, where $G(\bK)$ is a norm (or sum of norms) for sparsity-promoting operation, the Euclidean projection operation at $\bK^{o}$ onto the $G(\bK)$ ball, i.e., $G(\bK) \leq s$, denoted by $\cP_G(\bK^{o};s)$,  is conducted by solving the following optimization problem:
\par\noindent\small
\begin{align}\label{eq:proj}
	\cP_G(\bK^{o};s) =\argmin_{\bK}\;  \frac{1}{2}\| \bK - \bK^{o} \|_F^2,\;\;\text{s.t.}\;\;G(\bK) \leq s.  
\end{align}
\normalsize
If $\bK^{o}$ is inside of the $G(\bK)$ ball, then the projection value is just $\bK^{o}$. Therefore, we assume that $\bK^{o}$ is outside of the ball, i.e., $G(\bK^{o}) > s$.

\subsubsection{Projection onto the $\ell_0$-ball} \label{subsubsec:proj_l0}
When $G(\bK) = \| \bK \|_0$, where $\ell_0$ norm of the matrix $\bK$ represents the number of non-zero elements in $\bK$, the projection onto the $\ell_0$-ball is simply conducted by taking $s$ largest elements in $\bK^{o}$ and setting other elements to zero to have the minimum objective value in \eqref{eq:proj}. The similar stratergy is used in the GraSP algorithm for sparse optimal control design introduced in \cite{lian2017game}. However, the major difference between the GraSP and the ISPA with projection onto the $\ell_0$-ball is that the GraSP updates the non-zero element support set by merging the sets at each iteration, while the ISPA does not merge the support index set at each iteration. This difference can cause the performance of sparsity in the feedback matrix $\bK$. The comparison results against the GraSP \cite{lian2017game} is presented in the numerical experiment section. Procedure \ref{alg:proj_l0} describes the detail steps of the projection onto the $\ell_0$-ball.

\begin{procedure}[t]
\caption{Z(). Projection onto the $\ell_0$-ball}
  \label{alg:proj_l0}
  \SetAlgoLined
{\small
   \KwIn{ $\bK^{o} \in \R^{m \times n}$, and $s > 0$ }
     Step 1: $U^{o}_{l,k}$ $\leftarrow$ $|K^{o}_{l,k}|$, $\forall l,k$ \par 
   	Step 2: Sort $\vect(\bU^{o})$ in a descending order: $\vect(\bU^{o})_{(1)} \geq \vect(\bU^{o})_{(2)} \geq ... \geq \vect(\bU^{o})_{(mn)}$ \par
   	Step 3: Take the first $s$ elements in the descending order and set rest of the elements to zero  \par
     Step 3:  $\cP_{\ell_0}(\bK^{o};s)_{l,k}$ $\leftarrow$ $ \sgn(K^{o}_{l,k}) \cdot U^{o}_{l,k}$, $\forall l,k$\par
      \KwOut{ $\cP_{\ell_0}(\bK^{o};s)$ }
}%
\end{procedure}

\subsubsection{Projection onto the $\ell_1$-ball} \label{subsubsec:proj_l1}
Let us denote the vectorization of the matrix $\bK$ as $\vect(\bK)$ by stacking the column vectors of $\bK$, the index set for the non-zero elements as $\cM$, and the sorted elements in a descending order of magnitude as $\vect(\bK)_{(i)}$. From these notations, $\vect(\bK)_{i}$, $i \in \cM$, represent the non-zero elements. And in the sorted vector, the non-zero elements are expressed as $\vect(\bK)_{(i)}$, $i = 1,2,..., | \cM |$, where $|\cM|$ is the cardinality of the index set $\cM$, and $|\vect(\bK)_{(1)}| \geq | \vect(\bK)_{(2)}| \geq ... \geq |\vect(\bK)_{( |\cM |)}| $.

Then, from \cite{duchi2008efficient}, the  projection onto the $\ell_1$-ball can be conducted as follows:
\par\noindent\small
\begin{align}
	\cP_{\ell_1}(\bK^{o};s)_{l,k} = \sgn(K^{o} _{l,k}) \cdot \max\{ |K^{o}_{l,k}|  - \lambda, 0 \},
\end{align}
\normalsize
where $\lambda$ is stated as follows:
\par\noindent\small
\begin{align}\label{eq:optimal_lambda}
	\lambda = \frac{1}{|\cM|} \bigg( \sum_{i=1}^{|\cM|} |\vect(\bK^{o})_{(i)}|- s \bigg).
\end{align}
\normalsize
The cardinality of the index set $\cM$, denoted by $|\cM|$, which represents the number of non-zero elements, can be obtained from the following lemma:
\begin{lemma}\cite[Lemma 2]{duchi2008efficient} \label{lemma:num_nnz}
Let $|\vect(\bK^{o})_{(i)}|$, $i \in \Omega := \{1,...,mn\}$, be the sorted elements of $\bK^{o}$ in a descending order of magnitude. Then, the number of non-zero elements after projection onto $\ell_1$-ball is stated as follows:
\par\noindent\small
\begin{align}
	| \cM | \hspace{-0.1em}= \hspace{-0.1em}\max \bigg\{ j \in \Omega \bigg| | \vect(\bK^{o})_{(j)}| \hspace{-0.1em}- \hspace{-0.1em}\frac{1}{j} \bigg(\hspace{-0.2em} \sum_{i=1}^j | \vect(\bK^{o})_{(i)} |\hspace{-0.1em} - \hspace{-0.1em}s \hspace{-0.2em}\bigg) \hspace{-0.2em}> \hspace{-0.2em}0 \bigg\} \label{eq:num_nnz}
\end{align}
\normalsize
\end{lemma}
\noindent Intuitively, \eqref{eq:num_nnz} can be understood as the operation finding the minimum $|\vect(\bK^{o})_{(j)}|$ among the sorted values $|\vect(\bK^{o})_{(j)}|$, $j=1,...,mn$, satisfying $|\vect(\bK^{o})_{(j)} |- \lambda > 0$ by varying $j$. Procedure \ref{alg:proj_l1} introduces the summary of the projection onto the $\ell_1$-ball from $\bK^{o}$. 

\begin{procedure}[t]
\caption{S(). Projection onto the $\ell_1$-ball}
  \label{alg:proj_l1}
  \SetAlgoLined
{\small
   \KwIn{ $\bK^{o} \in \R^{m \times n}$, and $s > 0$ }
     Step 1: $U^{o}_{l,k}$ $\leftarrow$ $|K^{o}_{l,k}|$, $\forall l,k$ \par 
   	Step 2: Sort $\vect(\bU^{o})$ in a descending order: $\vect(\bU^{o})_{(1)} \geq \vect(\bU^{o})_{(2)} \geq ... \geq \vect(\bU^{o})_{(mn)}$ \par
     Step 3: Compute the number of non-zero elements $| \cM |$ via \eqref{eq:num_nnz} with the sorted $\vect(\bU^{o})$\par
     Step 4: Compute $\lambda$ via \eqref{eq:optimal_lambda} with the sorted $\vect(\bU^{o})$ and $| \cM |$ \par
     Step 5:  $\cP_{\ell_1}(\bK^{o};s)_{l,k}$ $\leftarrow$ $ \sgn(K^{o}_{l,k}) \cdot \max\{ U^{o}_{l,k} - \lambda, 0\}$, $\forall l,k$\par
      \KwOut{ $\cP_{\ell_1}(\bK^{o};s)$ }
}%
\end{procedure}

\subsubsection{Projection onto the block sparsity-ball}
Since the state variables can be grouped into each agent as shown in Fig. \ref{fig:DS}, in order to reduce the communication links among multi-agents, the block sparsity can be considered for a possible approach. For the block sparsity operation, we define the sparsity-promoting function $G(\bK) = \sum_{i,j} \| \bK_{i,j} \|_F$, where $\bK_{i,j}$ is the $i$-row and the $j$-th column block of $\bK$. Then, the projection onto the block sparsity ball, denoted by $\cP_{bs}(\bK^{o};s)$, is stated as follows:
\par\noindent\small
\begin{align}\label{eq:proj_bs}
	\cP_{bs}(\bK^{o};s) =\argmin_{\bK}\;  \frac{1}{2}\| \bK - \bK^{o} \|_F^2,\;\;\text{s.t.}\;\;\sum_{i,j} \| \bK_{i,j} \|_F \leq s.  
\end{align}
\normalsize
In order to have the close-form solution to \eqref{eq:proj_bs}, we denote the variable $\bK$ by the summation of the same direction of $\bK^{o}$ part and its orthogonal part, as follows:
\par\noindent\small
\begin{align}\label{eq:bs_proj_inital}
	\bK_{i,j} = t_{i,j} \frac{\bK^{o}_{i,j}}{\| \bK^{o}_{i,j}\|_F} + \bZ_{i,j},
\end{align} 
\normalsize
where $t_{i,j}$ is a scalar variable for the $i$-th row and $j$-th column block, and $\bZ_{i,j}$ is an orthogonal matrix to $\bK^{o}_{i,j}$, i.e., $\langle \bK_{i,j}^{o}, \bZ_{i,j} \rangle = 0$. Then, \eqref{eq:proj_bs} can be restated as follows:
\par\noindent\small
\begin{align*}
	& \underset{\bt,\; \bZ}{\text{minimize}}\;  \frac{1}{2} \sum_{i,j} \bigg\|  t_{i,j} \frac{\bK^{o}_{i,j}}{\| \bK^{o}_{i,j}\|_F} + \bZ_{i,j} - \bK^{o}_{i,j} \bigg\|_F^2 \\
	& \quad\quad\text{s.t.}\;\;\sum_{i,j} \bigg\|  t_{i,j} \frac{\bK^{o}_{i,j}}{\| \bK^{o}_{i,j} \|_F} + \bZ_{i,j} \bigg\|_F \leq s,
\end{align*}
\normalsize
which implies 
\par\noindent\small
\begin{align}\label{eq:proj_bs_linear}
		 & \underset{\bt,\; \bZ}{\text{minimize}}\;  \frac{1}{2} \sum_{i,j}  \bigg[\bigg( t_{i,j} - \| \bK^{o}_{i,j}\|_F \bigg)^2 + \| \bZ_{i,j} \|_F^2 \bigg]\nonumber \\
		 &\quad\quad \text{s.t.}\;\;\sum_{i,j} \sqrt{  t_{i,j}^2 + \| \bZ_{i,j}  \|^2_F } \leq s.
\end{align}
\normalsize
In order for the objective function of \eqref{eq:proj_bs_linear} to be minimum, $\bZ_{i,j}$ needs to be zero. Therefore, we have 
\par\noindent\small
\begin{align}
		 \underset{\bt}{\text{minimize}}\;  \frac{1}{2} \sum_{i,j}  \bigg( t_{i,j} - \| \bK^{o}_{i,j}\|_F \bigg)^2,\;\;\text{s.t.}\;\;\sum_{i,j} |  t_{i,j}|  \leq s.
\end{align}
\normalsize
Additionally, since $\| \bK^{o}_{i,j} \|_F \geq 0$, $t_{i,j}$ will be chosen to be positive. Hence, we simply have
\par\noindent\small
\begin{align}
		 & \underset{\bt}{\text{minimize}}\;  \frac{1}{2} \sum_{i,j}  \bigg( t_{i,j} - \| \bK^{o}_{i,j}\|_F \bigg)^2 \nonumber \\
		 & \quad\quad\;\; \text{s.t.}\;\;\sum_{i,j}  t_{i,j}  \leq s,\;\; t_{i,j} \geq 0, \forall i,j,
\end{align}
\normalsize
which is the projection onto the positive simplex with the assumption that $\bK^{o}$ is outside of the ball, i.e., $\sum_{i,j} \| \bK^{o}_{i,j} \|_F > s$. Note that this optimization problem is the same as the projection onto $\ell_1$ ball, but the elements are all positive. Hence, the optimal $t_{i,j}^{\star} = \max\{\| \bK^{o}_{i,j} \|_F - \lambda, 0 \}$, where $\lambda$ is obtained from \eqref{eq:optimal_lambda} and \eqref{eq:num_nnz}. Therefore, by plugging the optimal solution $\bt^{\star}$ and $\bZ^{\star}$ to \eqref{eq:bs_proj_inital}, we have the following result for the projection onto the block sparsity ball:
\par\noindent\small
\begin{align}\label{eq:proj_bs_final}
	\cP_{bs}(\bK^{o};s)_{i,j}  = \max \bigg\{ \| \bK^{o}_{i,j} \|_F -  \lambda, 0 \bigg\} \cdot \frac{\bK^{o}_{i,j} }{\| \bK^{o}_{i,j} \|_F}.
\end{align}
\normalsize
Procedure \ref{alg:proj_bs} summarizes the projection steps onto the block sparsity-ball.

\begin{procedure}[t]
\caption{B(). Projection onto the block sparsity-ball}
  \label{alg:proj_bs}
  \SetAlgoLined
{\small
   \KwIn{ $\bK^{o} \in \R^{m \times n}$, and $s > 0$ }
     Step 1: $U^{o}_{i,j}$ $\leftarrow$ $\| \bK^{o}_{i,j}\|_F$, $\forall i,j$, where $\bK^{o}_{i,j}$ is the $i$-th row and $j$-th column block matrix of $\bK^{o}$ \par 
   	Step 2: Sort $\vect(\bU^{o})$ in a descending order: $\vect(\bU^{o})_{(1)} \geq \vect(\bU^{o})_{(2)} \geq ... \geq \vect(\bU^{o})_{(p)}$, where $p$ is the number of blocks in $\bK^{o}$\par
     Step 3: Compute the number of non-zero elements $| \cM |$ via \eqref{eq:num_nnz} with the sorted $\vect(\bU^{o})$\par
     Step 4: Compute $\lambda$ via \eqref{eq:optimal_lambda} with the sorted $\vect(\bU^{o})$ and $| \cM |$ \par
     Step 5: Compute \eqref{eq:proj_bs_final} to obtain $\cP_{bs}(\bK^{o};s)_{i,j}$, $\forall i,j$ \par
      \KwOut{ $\cP_{bs}(\bK^{o};s)$ }
}%
\end{procedure}

\subsection{Iterative Sparse Projection Algorithm} \label{subsec:ISPA}
With the projection operations introduced in the previous sub-section, we introduce the ISPA for sparse LQR optimal control design in this subsection. The ISPA is organized with three steps:  gradient descent, projection onto  a sparsity-promoting ball, and stability check. Since the initial point of the ISPA is the solution obtained by solving LQR problem without sparsity-promoting function constraint, it has the minimum $J(\bK)$ value but normally $G(\bK) > s$. Hence, the first step of the ISPA is finding a feasible solution to \eqref{eq:LQR_prob_change}. And then, from the feasible solution, the ISPA conducts gradient descent and projection onto a sparsity-promoting ball, e.g., $\ell_1$-ball, iteratively until the Euclidean distance between two adjacent-estimated feedback matrice is less than a certain predetermined small tolerant value $\epsilon$. Since the estimated solution $\hat{\bK}$ is needed to be in $\cF$, after the projection operation, we check the stability of the system by computing the largest eigenvalue of $\bA-\bB \bK^{(t+1)}$, where $\bK^{(t+1)}$ is the projection result at $t$-th iteration. If the estimated feedback matrix $\bK^{(t+1)}$ is not in $\cF$, then we do the backtracking operation with a smaller step-size. Algorithm \ref{alg:ISPA} introduces the summary of the algorithm steps, where the projection onto a sparsity-promoting ball follows the steps in Procedure \ref{alg:proj_l0}, \ref{alg:proj_l1}, or \ref{alg:proj_bs}.

The ISTA and the ISPA have similar algorithm steps. The major difference of the ISTA and the ISPA is that the ISTA has thresholding operation, while the ISPA has projection operation, which is the result from the difference of the optimization problems between \eqref{eq:LQR_prob} and \eqref{eq:LQR_prob_change}. The benefit of the ISPA compared to the ISTA is that in the ISPA, the step-size is the only major parameter, which can be simply adjusted in the algorithm, while the ISTA has parameters for step-size as well as thresholding level which are connected each other with $\rho$ as shown in \eqref{eq:update_l1}, \eqref{eq:update_blkl1}, \eqref{eq:update_wl1},  and \eqref{eq:update_blkw1}.

\begin{algorithm}[t]
  \caption{Iterative Sparse Projection Algorithm (ISPA) for Sparse LQR Optimal Control Design}
  \label{alg:ISPA}
  \SetAlgoLined
  \SetKwRepeat{Do}{do}{while}%
{\small
   \KwIn{ $\bA$, $\bB$, $\bS_x$, $\bQ \succeq 0$, $\bR \succ 0$, $s_{\gamma}$, $\alpha < 1$, $\epsilon$, $MaxItr$, $\rho_0$  }
   \textbf{Initialization}: $\bK^{(0)} \leftarrow$ a stable dense matrix $\bK \in \cF$ from a solution to LQR problem  \par
   \For { $t=0$  \KwTo $MaxItr$ }
   {
   		   \uIf {$ t = 0$}
   		   {
   		   		$\bK^{(1)}$ $\leftarrow$ finding a feasible solution of \eqref{eq:LQR_prob_change} from $\bK^{(0)}$\par
   		   }\Else{
   		   $\bP^{(t)}$ $\leftarrow$ solution of \eqref{eq:lya_P} from $\bK^{(t)}$ \par
   		   $\bL^{(t)}$ $\leftarrow$ solution of \eqref{eq:lya_L} from $\bK^{(t)}$ \par
   		   $\grad J(\bK^{(t)}) $ $\leftarrow$ calculating the gradient value of $J(\bK)$ at $\bK^{(t)}$ from \eqref{eq:grad_J}  \par
   		   \Do{ $\rho^{(t)} \neq \rho_{0}$ }{
	             $\tilde{\bK}$ $\leftarrow$ $\bK^{(t)} - \rho^{(t)} \grad J(\bK^{(t)})$   \Comment{Gradient descent} \par
	             $\bK^{(t+1)}$ $\leftarrow$ $\cP_G(\tilde{\bK};s_{\gamma})$ \Comment{Projection onto the sparsity function $G(\cdot)$ ball} \par	             
		   			\uIf { $\max(\Re(\lambda(\bA - \bB\bK^{(t+1)})) \geq 0$ \textcolor[rgb]{1,0,0}{or} $J(\bK^{(t+1)}) > J(\bK^{(t)}) + \beta \| \grad J(\bK^{(t)})\|$}    
		   			{
		   				$\rho^{(t)}$ $\leftarrow$ $\rho^{(t)} / \alpha$ \Comment{Backtracking with a smaller step-size} \par
		   			}\Else {
						$\rho^{(t)}$ $\leftarrow$ $\rho_{0}$ \par		
					}
			}
			\If { $\| \bK^{(t)} - \bK^{(t-1)} \|_F < \epsilon$ } 
			{
				break
			}
			}			
}
  \KwOut{ Structured feedback matrix $\bK$ }
 }
\end{algorithm}



The convergence speed of the ISTA and the ISPA in practical situations are highly related to the step-size of the gradient descent in the algorithms.  An optimal step-size can be different depending on given problems and situations. However, for a given system, we can find a good step-size which can provide much fast convergence speed by using data-driven approach. In order to improve the convergence speed of the ISTA and the ISPA, let us  introduce the model-based DNNs in the next section.

\section{Model-based Deep Neural Networks for Sparse LQR Optimal Control Design}
\label{sec:DNN}
In large-scale distributed control systems, it is sometimes necessary to quickly find stable feedback matrices for dynamic system models, due to variations in system states or malfunctions. However, existing algorithms such as ISTA, ISPA, and ADMM \cite{lin2013design} can be computationally burdensome, making it difficult to efficiently obtain sparse optimal feedback matrices within a limited time frame for these systems. In such cases, DNNs can serve as promising alternatives, offering reduced computation time while providing sparse optimal feedback matrices. Furthermore, DNNs can also offer suitable initial values for warm-starting the algorithms.

The training of DNNs is performed offline using a dataset that includes state and input matrices. This dataset is generated by introducing variations $(\Delta \bA, \Delta \bB)$ to a given set of ground-truth state and input matrices $(\bA^{o}, \bB^{o})$. Specifically, the dataset, comprising $r$ data points, can be expressed as $\{ (\bA^{o} + \Delta \bA^i, \bB^{o} + \Delta \bB^{i}, \bS_x^{i}) \}_{i=1}^{r}$, where $\Delta \bA^i$, $\Delta \bB^{i}$, and $\bS_x$ represent small variations at the $i$-th data point for the state matrix and the input matrix, and different initial inputs.

\subsection{Implementation of DNN}
We utilize the ISTA, introduced in Section \ref{sec:algo}, as a baseline for the architecture of a DNN in sparse optimal control design. Thus, a model-based method serves as a bridge between a general DNN and an iterative algorithm. Since an iterative algorithm involves repeating the same operations multiple times, the neural network built based on the iterative algorithm can be interpreted as a DNN, with each layer representing one iteration of the algorithm. 

The model-based DNN for sparse optimal control design aim to enhance the accuracy and computational efficiency of the proposed algorithm even by reducing the number of layers, and thus, the number of iterations in the ISTA. This reduction is achieved by training the DNN to determine the optimal step size $1/\rho^{(t+1)}$ and thresholding level $\gamma/\rho^{(t+1)}$.

\begin{figure}[t]
    \centering
    \includegraphics[scale=0.30]{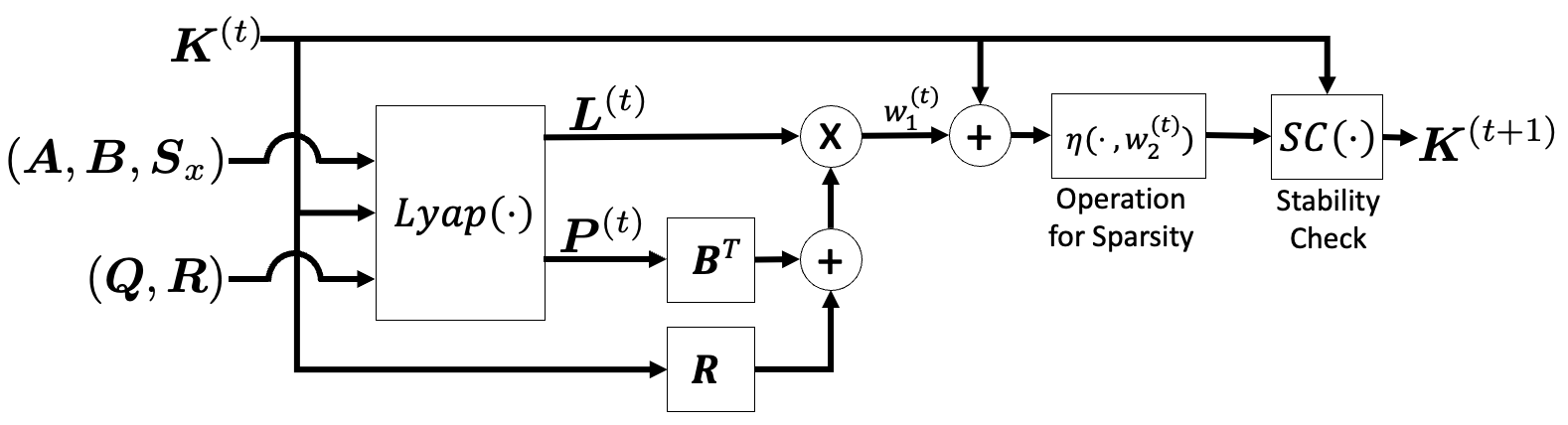}
    \caption{\small Data flow of the $t$-th layer of the DNN based on the ISTA (resp. the ISPA) introduced in Algorithm \ref{alg:ISTA} (resp. Algorithm \ref{alg:ISPA}), where $w^{(t)}_1$ and $w^{(t)}_2$ are the training variables at the $t$-th layer of the DNN. For the ISPA, the operation for sparsity block is understood as the projection operation.}
    \label{fig:dataflow_ITA_RNN}
\end{figure}

Fig. \ref{fig:dataflow_ITA_RNN} illustrates the data flow of the $t$-th layer of the DNN based on the ISTA. In this flow, $w^{(t)}_1$ and $w^{(t)}_2$ are the training variables in the $t$-th layer, corresponding to the step size and threshold level, respectively. The Lyapunov solver block, denoted as $Lyap(\cdot)$, solves the Lyapunov equations \eqref{eq:lya_P} and \eqref{eq:lya_L}, while $\eta(\cdot; \cdot)$ denotes a sparsity-promoting operation. To construct the DNN, we concatenate $l$ blocks, where each block corresponds to the data flow shown in Fig. \ref{fig:dataflow_ITA_RNN} and represents a layer. This unfolded model-based DNN, comprising $l$ layers, is referred to as DNN based ISTA, or simply DNN-ISTA. It is worth noting that while there is a final solution $\bK^{(l)}$, we can also access intermediate solutions ranging from $\bK^{(1)}$ to $\bK^{(l-1)}$, all of which lie within the feasible set $\cF$.


\begin{figure}[t]
    \centering
    \includegraphics[scale=0.24]{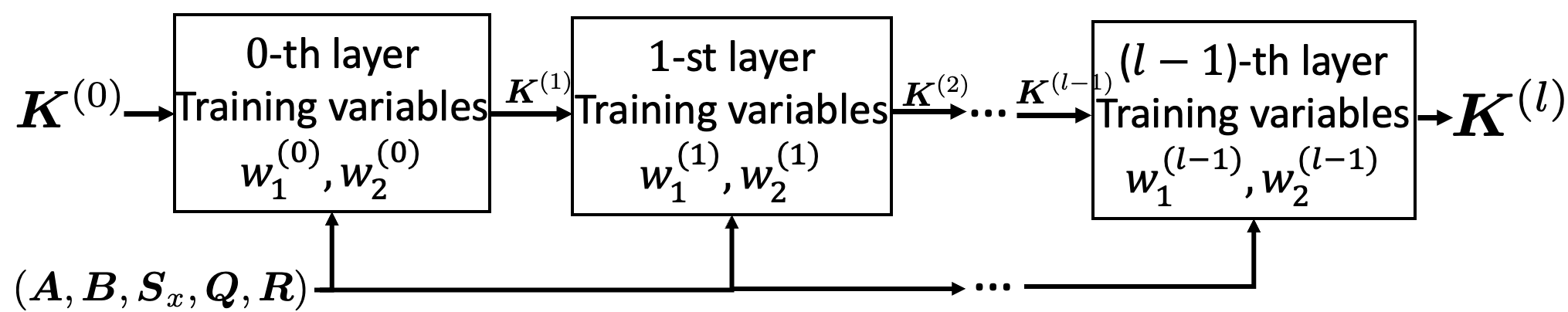}
    \caption{\small Illustration of the unfolded DNN based on ISTA (DNN-ISTA) for the sparse optimal control design with $l$ layers.}
    \label{fig:dataflow_ITA_RNN_BIGPIC}
\end{figure}

\subsection{FISTA-based DNN}
\label{subsec:modi_algo}
To further enhance the convergence speed of the ISTA, we can employ the momentum method and Nesterov accelerated gradient update \cite{nesterov2003introductory}. Drawing inspiration from the Fast ISTA (FISTA) concept \cite{beck2009fast,daubechies2004iterative}, we propose a modification by combining the current and previous results using a parameter $w_3$, as expressed by:
\par\noindent\small
\begin{align}\label{eq:modi_K}
	 \tilde{\bK}^{(t+1)} = w_3 \bK^{(t+1)} + (1-w_3) \bK^{(t)}.
\end{align}
\normalsize
Here, $\tilde{\bK}^{(t+1)}$ represents the input to the next layer. If $w_3=1$, then it becomes the same as the standard ISTA. However, in the context of sparse optimal control design, the conventional FISTA updating scheme for the parameter $w_3$ (equivalently $\beta^{(t)}$ in \eqref{eq:FISTA_step}) cannot be directly applied due to the constraints imposed by the feasible set $\cF$. Nevertheless, through a data-driven approach, we can learn the parameter $w_3$ during the training process.

\begin{figure}[t]
    \centering
    \includegraphics[scale=0.23]{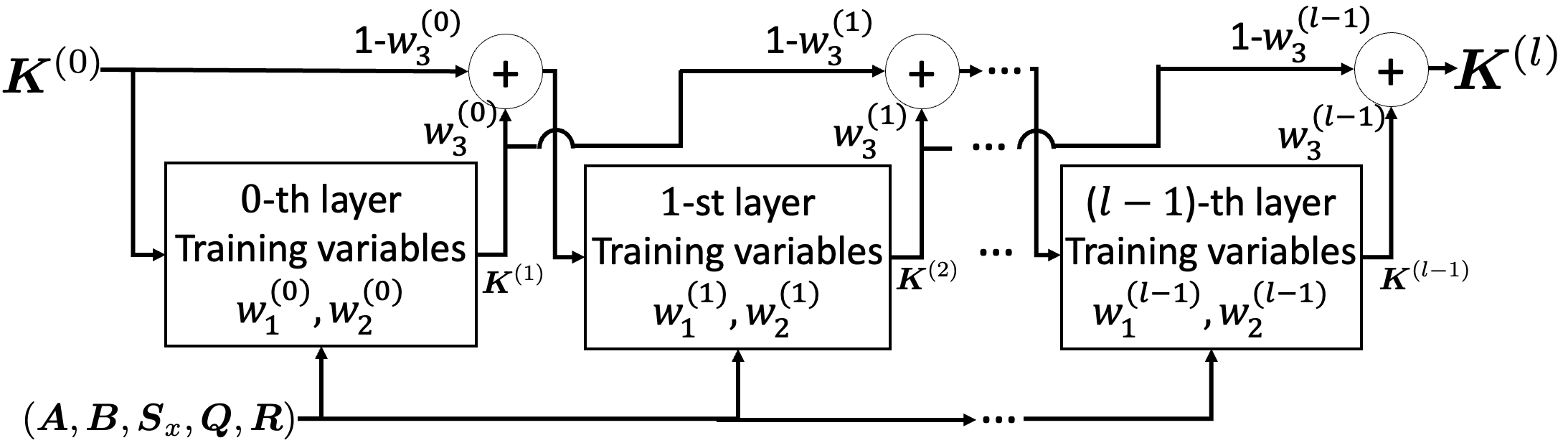}
    \caption{\small Illustration of the unfolded FISTA-based DNN (DNN-FISTA) for the sparse optimal control design with $l$ layers.}
    \label{fig:dataflow_FISTA_DNN_BIGPIC} 
    \vspace{-1em}
\end{figure}

Fig. \ref{fig:dataflow_FISTA_DNN_BIGPIC} depicts the implementation of the DNN based on FISTA. While the parameter $w_3$ originates from the FISTA updating scheme, it can also be viewed as a jumping connection between adjacent intermediate results in a general DNN. This concept has been explored in various DNN designs \cite{han2015deep,amos2017input,chen2018optimal} to reduce the complexity of the network and mitigate the issue of gradient vanishing during back-propagation \cite{pascanu2013difficulty,glorot2010understanding}.

\subsection{Sparsity-promoting operation in DNNs} \label{subsec:sparse_funcs}
In the ISTA, two types of sparsity-promoting functions are introduced: the shrinkage-thresholding operator $\cS(\cdot)$ and the block thresholding operator $\cB(\cdot)$. In \eqref{eq:update_l1} and \eqref{eq:update_blkl1}, the step size $\frac{1}{\rho^{(t+1)}}$ and the thresholding level $\frac{\gamma}{\rho^{(t+1)}}$ are connected through $\rho^{(t+1)}$. However, when training the DNNs, we can treat these parameters as separate training variables to provide the DNNs with greater flexibility.

For the shrinkage-thresholding operator $\cS(\cdot)$, we introduce a training variable $w$ to learn the thresholding level, resulting in the following formulation:
\par\noindent\small
\begin{align}\label{eq:shrinkage_nn}
	\cS( \bK; w )_{l,k} := \sgn( K_{l,k}) \max\{ | K_{l,k} | - w, 0\}.
\end{align}
\normalsize
Here, $w$ represents the training variable. Similarly, for the block shrinkage-thresholding operator $\cB(\cdot)$, we can incorporate a training variable, leading to:
\par\noindent\small
\begin{align}
	\cB( \bK; w )_{i,j} := \max\big\{  \| \bK_{i,j} \|_F - w, \;0 \big\} \cdot \frac{\bK_{i,j}}{\| \bK_{i,j} \|_F}.
\end{align}
\normalsize

Moreover, we can consider training the parameters for each layer independently. As a result, in a DNN comprising $l$ layers, we would have $O(l)$ training variables.

\subsection{Stability check and training DNNs}
In Algorithm \ref{alg:ISTA}, the process of finding a sparse matrix $\bK^{(t+1)} \in \cF$ involves a backtracking operation if the $(t+1)$-th intermediate matrix $\bK^{(t+1)}$ is not in $\cF$. However, implementing the backtracking operation in a DNN can be challenging as it requires an inner loop, which is the do-while loop in Algorithm \ref{alg:ISTA}. Moreover, during DNN training, it is possible to determine an optimal step size and threshold level for each layer. Consequently, we simplify the backtracking operation to a check for whether the intermediate feedback matrix $\bK^{(t+1)}$ belongs to $\cF$ or not. If $\bK^{(t+1)} \in \cF$, we use it as the input for the next layer of the DNN. Otherwise, we assign $\bK^{(t)} \in \cF$ to $\bK^{(t+1)}$. This approach allows us to obtain an intermediate feedback matrix $\bK^{(t+1)}$ that can asymptotically stabilize the distributed system. \textcolor[rgb]{0,0,0}{To guarantee that $\bK^{(t+1)} \in \cF$, we need to use a stable initial feedback matrix $\bK^{(0)}$, i.e., $\bK^{(0)} \in \cF$.}

The training of DNNs involves solving the following loss minimization problem with respect to the training parameters:
\par\noindent\small
\begin{align}\label{eq:training}
	& \underset{ \{w^{(t)}_1, w_2^{(t)}, w_3^{(t)} \}_{t=0}^{l-1}}{\text{minimize}} \; \sum_{i=1}^r \| {\bK^i}^{\star} - DNN_{[0:l-1]}(\bK^{(0)}) \|_F^2,
\end{align}
\normalsize
where $DNN_{[0:l-1]}(\cdot)$ represents the output of the DNN model based on FISTA with $l$ layers, as depicted in Fig. \ref{fig:dataflow_FISTA_DNN_BIGPIC}. \textcolor[rgb]{0,0,0}{${\bK^i}^{\star}$ represents the $i$-th target sparse feedback matrix corresponding to the $i$-th perturbed model, i.e., $(\bA^{o} + \Delta \bA^i, \bB^{o} + \Delta \bB^{i})$.} The training variables $\{ w^{(t)}_1, w_2^{(t)}, w_3^{(t)} \}_{t=0}^{l-1}$ are adjusted using back-propagation with the provided dataset during DNN training.

\textcolor[rgb]{0,0,0}{\textbf{Remark:} The ISPA shares similar algorithmic steps with the ISTA, but includes an additional operation to calculate the thresholding level parameter $\lambda$ in \eqref{eq:optimal_lambda} and \eqref{eq:num_nnz} based on the number of non-zero elements $s$. Since both the thresholding level $\lambda$ and the step size can be learned from the dataset, the DNN based on the ISPA is essentially the same as the DNN-ISTA. \textcolor[rgb]{0,0,0}{Additionally, the model-based DNNs discussed in this section can be viewed as an alternative method for implementing the proposed algorithms, namely ISTA and ISPA.}
}

\section{Numerical experiments}
\label{sec:experiment}
In the numerical experiments, we consider various system models including a distributed multi-agent control system model and 16 machine system model with 86 transmission lines and 68 buses extracted from the GE report \cite{general1983singular}. We firstly compare the performance of the ISTA and the ADMM \cite{lin2013design} which solve the optimization problem \eqref{eq:LQR_prob} in the aspects of sparsity level, LQR cost, and execution time. With different types of regularization terms for $G(\bK)$ introduced in Section \ref{subsec:algo2}, we conduct the numerical experiments by varying $\gamma$ from 0.1 to 5, and compare the results obtained from the ISTA and the ADMM \cite{lin2013design}. Through the whole simulations, we set $\bQ$ and $\bR$ to the identity matrices. For the initial point of the both algorithms, we calculate the Linear-Quadratic Regulator (LQR) function in Matlab given matrices $\bA$, $\bB$, $\bQ$, and $\bR$, which normally provides a dense feedback matrix $\bK$. We also compare the ISPA and the GraSP \cite{lian2017game}. After the performance comparison between the algorithms, we demonstrate the performance improvement by using the model-based DNNs introduced in Figs. \ref{fig:dataflow_ITA_RNN_BIGPIC} and \ref{fig:dataflow_FISTA_DNN_BIGPIC}. All numerical experiments are performed in Macbook pro 2016 with Intel Core i7 and 16 GB RAM. 

\begin{figure*}[t]
    \centering
    \subfloat[Sparsity]{\includegraphics[scale=0.32]{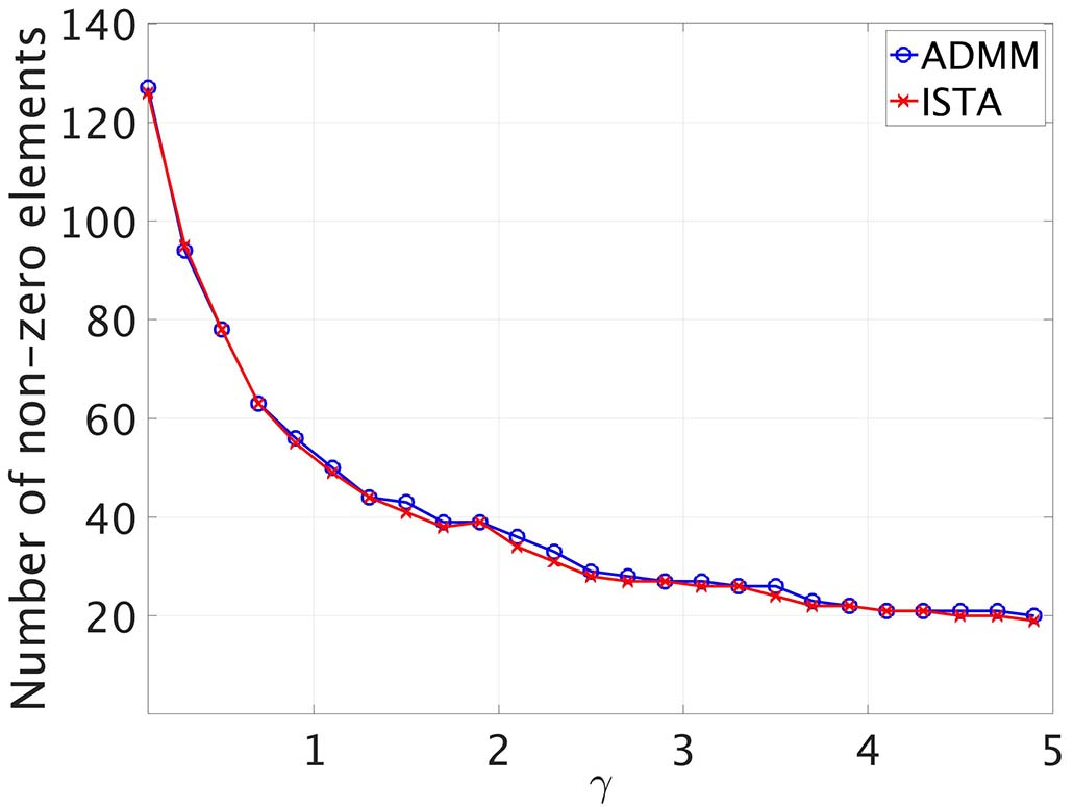}}
    \subfloat[$J(\bK)$ value]{\includegraphics[scale=0.32]{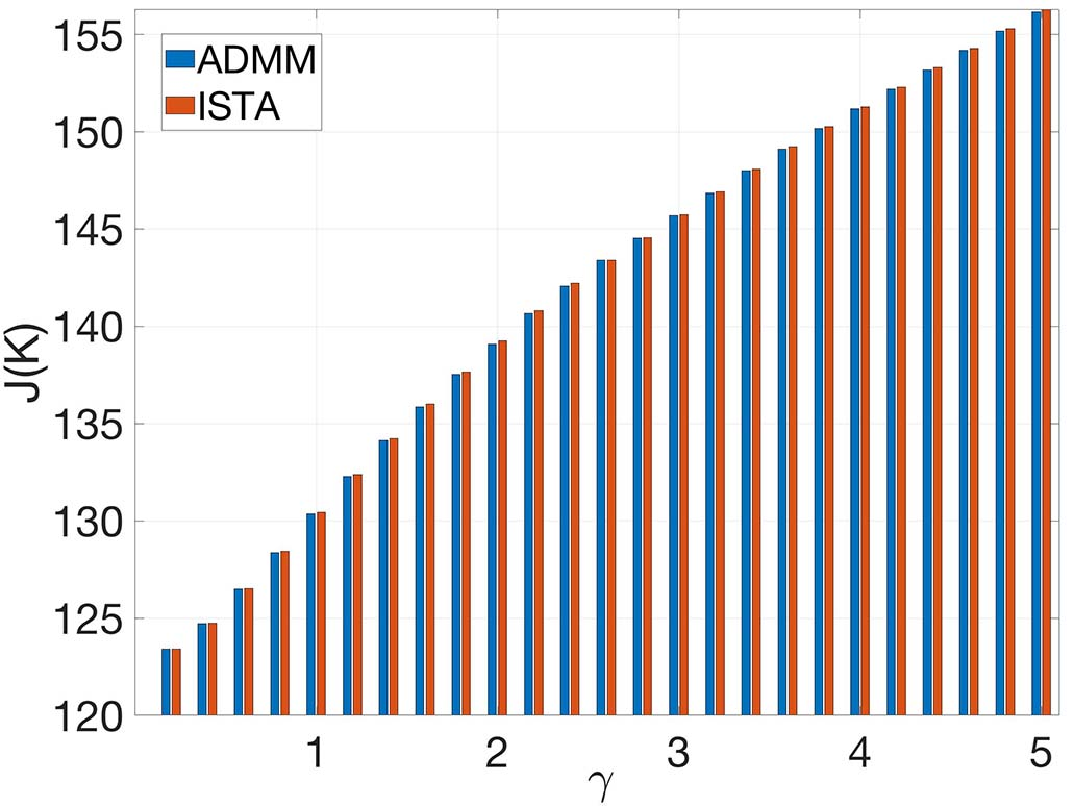}} 
    \subfloat[Complexity]{\includegraphics[scale=0.32]{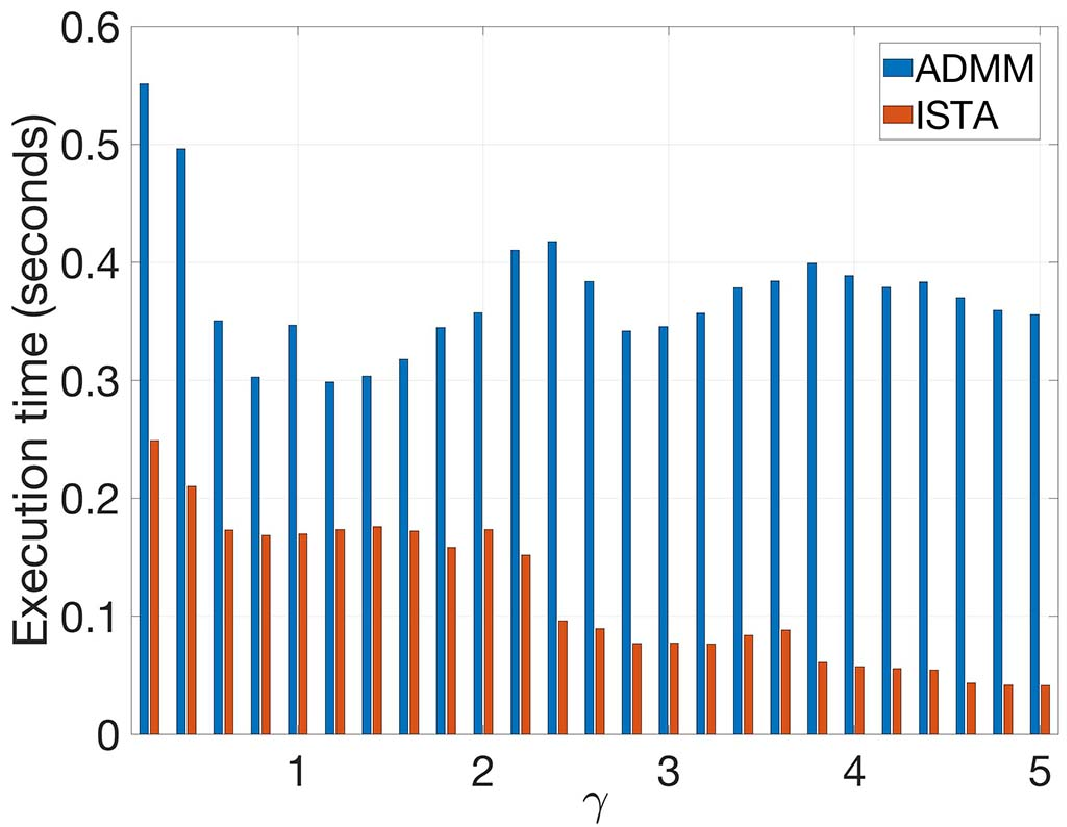}}        
    \caption{Comparison between the ISTA and the ADMM in the distributed multi-agent control system model.  $G(\bK) = \| \bK \|_1$ is used. }
    \label{fig:comparsion_ADMM_ISTA_L1}
\end{figure*}

\begin{figure*}[t]
    \centering
    \subfloat[Sparsity]{\includegraphics[scale=0.64]{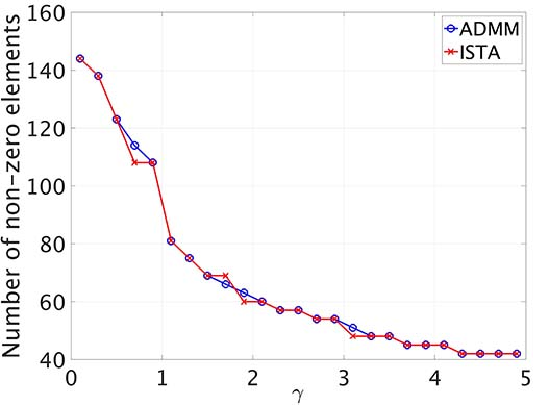}}
	\subfloat[$J(\bK)$ value]{\includegraphics[scale=0.64]{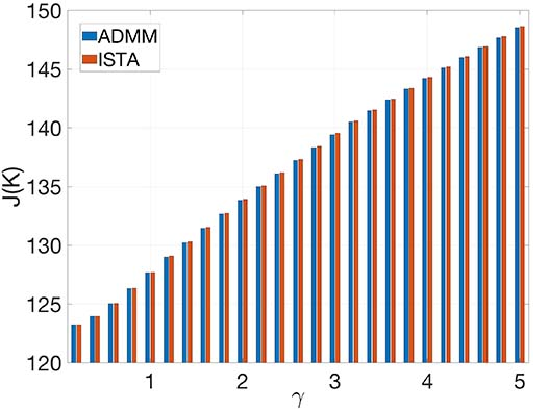}}  		
	\subfloat[Complexity]{\includegraphics[scale=0.64]{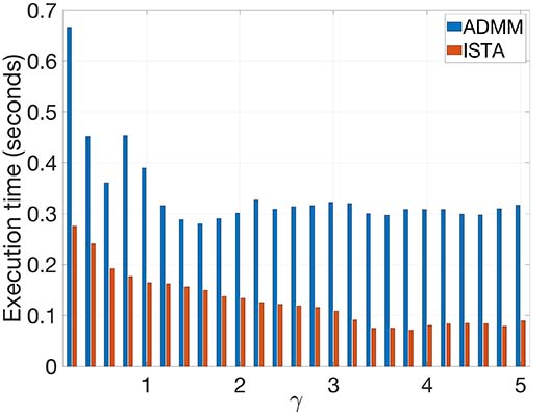}}	
    \caption{Comparison between the ISTA and the ADMM in the distributed multi-agent control system model. The block sparsity promoting function, i.e., $G(\bK) = \sum_{i,j} \| \bK_{i,j} \|_F$ is used. }
    \label{fig:comparsion_ADMM_ISTA_BlkL1}
\end{figure*}

\subsection{Distributed multi-agent control system model} 
\label{subsec:simulation_prob}
We address a distributed multi-agent control system model having $N$ multi-agents, where the $i$-th agent has the following dynamics:
\par\noindent\small
\begin{align}
	\dot{\bx}_i = \bA_{i,i} \bx_i - \frac{1}{2} \sum_{j=1}^N (i-j) (\bx_i - \bx_j) + {\bB}_{i,i} \bu_i, 
\end{align}
\normalsize
where $\bA_{i,j}$ represents the $i$-th row and the $j$-th column block of matrix $\bA$. ${\bA}_{i,i}$ and ${\bB}_{i,i}$ are given as follows:
\par\noindent\small
\begin{align*}
	\bA_{i,i} = \begin{bmatrix}-6 & 0 & -3 \\ 3 & -6 & 0 \\ 0 & 3 & -6 \end{bmatrix},\;
	{\bB}_{i,i} = \begin{bmatrix} 3 & 0 \\ 0 & 3  \\ 0 & 0\end{bmatrix}.
\end{align*}
\normalsize
\textcolor[rgb]{0,0,0}{By considering the expectation of a randomly generated initial input vector $\bx(0)$, we set $\bS_x$ to $3I$.} The overall size of the state matrix $\bA$ considering $N$ numbers of multi-agents is then $ 3N \times 3N$. The dimension of the feedback matrix is $2N \times 3N$. For the regularization term $G(\bK)$, we consider both $\| \bK \|_1$ and $\sum_{i,j} \| \bK_{i,j} \|_F$. Depending on the structure of the feedback matrix, we will have differnt connection among the $N$ numbers of multi-agents, and the agents will share their state information through the communication network. Similar problem models are considered in \cite{lin2013design, jovanovic2008passivity}.

\subsubsection{Comparison between ISTA and ADMM} \label{subsubsec:comp_admm_ista}
We conduct numerical experiments to compare the performance of the ISTA introduced in Algorithm \ref{alg:ISTA} against the ADMM method introduced in \cite{lin2013design} for the sparse LQR optimal control design problem \eqref{eq:LQR_prob} in the distributed multi-agent control system model. For tuning parameters in the ISTA, we set $\alpha=1.5$, $\epsilon=10^{-4}$, $MaxItr=10000$, and $\rho_0=100$. The parameters in the ADMM are set to $\rho=100$ in \eqref{eq:ADMM_step}, the tolerance for the absolute criterion $\epsilon_{abs}= 10^{-4}$, and the tolerance for the relative criterion $\epsilon_{rel}= 10^{-2}$ \cite{boyd2011distributed}. We vary $\gamma$ value from $0.1$ to $5$. The $N=5$ multi-agents in the system are considered for this comparison.

Fig. \ref{fig:comparsion_ADMM_ISTA_L1} shows the performance in sparsity level, LQR cost, i.e., $J(\bK)$ value at the estimated solution, and the execution time in seconds respectively when $G(\bK) = \| \bK \|_1$. Additionally, Fig. \ref{fig:comparsion_ADMM_ISTA_BlkL1} shows the comparison results when $G(\bK)$ is chosen for block sparsity, i.e., $G(\bK) = \sum_{i=1}^N \sum_{j=1}^N \| \bK_{i,j} \|_F$, where $\bK_{i,j} \in \R^{2 \times 3}$ is the $i$-th row and $j$-th column block of $\bK$. We compute the average execution time for the ISTA and the ADMM from 25 trials by varying $\gamma$ from 0.1 to 5, and obtain 0.3729 seconds for the ADMM and 0.1167 seconds for the ISTA. From the numerical experiments shown in Figs. \ref{fig:comparsion_ADMM_ISTA_L1} and \ref{fig:comparsion_ADMM_ISTA_BlkL1}, we demonstrate that the ISTA can have less computational complexity than the ADMM with similar sparsity performance and $J(\bK)$ value. Since in the ISTA, the cheap gradient descent step is conducted at each step, the number of iterations in the ISTA can be larger than that of the ADMM. However, the overall complexity of the ISTA can be reduced than that of the ADMM.

\begin{figure}[t]
    \centering
    \subfloat[Complexity]{\includegraphics[scale=0.49]{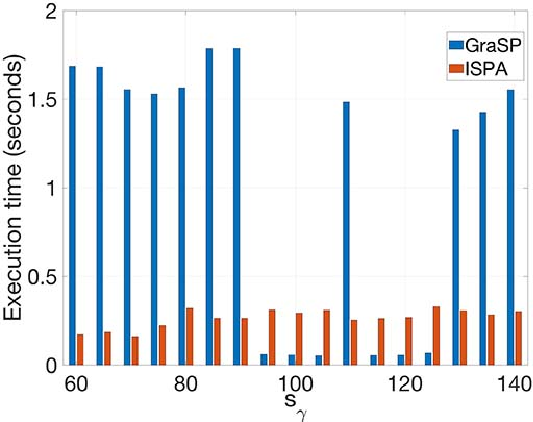}}
    \subfloat[$J(\bK)$ value]{\includegraphics[scale=0.49]{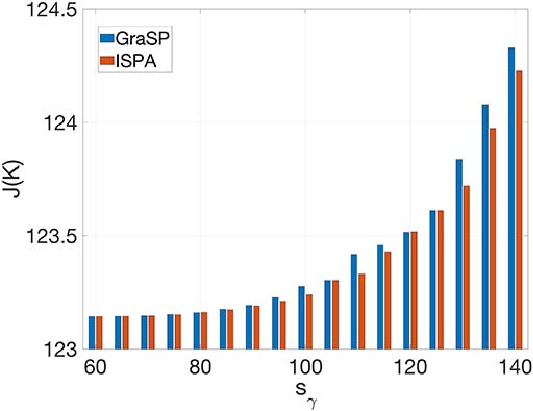}}    
    \caption{Comparison between the the ISPA with $G(\bK) = \| \bK \|_0$ and the GraSP in the distributed multi-agent control system model. (a) Complexity. (b) $J(\bK)$ value.}
    \label{fig:comparison_ISPA_GraSP}
\end{figure}

\subsubsection{Comparison between ISPA and GraSP}
We compare the ISPA and the GraSP method introduced in \cite{lian2017game}. Both algorithms are designed to solve \eqref{eq:LQR_prob_change}. Since the GraSP method takes the $s$ numbers of largest elements in the estimated feedback matrix  $\bK^{(t)}$ at the $t$-th iteration and sets the other elements zero, the GraSP can be understood as the one of the iterative $\ell_0$-ball projection algorithms. We compare the ISPA with $\ell_0$-ball against the GraSP method. In the $\ell_0$-ball, the number of non-zero elements in the feedback matrix $\bK$ is varied from 60 to 140 out of $150$ elements. The other tuning parameters in the ISPA are set to $\alpha=0.7$, $\epsilon=10^{-4}$, $MaxItr=10000$, and $\rho_0=1$. In the GraSP, we use the same stopping creteria as the ISPA; namely, when the Euclidean distance between two adjacent feedback matrices are less than the tolerance $\epsilon = 10^{-4}$, we stop the GraSP. The maximum iteration number, $MaxItr$, for the GraSP is set to $1000$, and the restricted Newton step of $J(\bK)$ is used  as proposed in\cite{lian2017game}. For both the ISPA and the GraSP, the LQR solution without the sparsity constraint is used as an initial point of the both algorithms. At each case of the sparsity level $s$, we measure the execution time and $J(\bK)$ value. 

Figs. \ref{fig:comparison_ISPA_GraSP} (a) and (b) show the execution time and the LQR cost. The average execution time for the GraSP is 1.0427 seconds, while it is 0.2655 seconds for the ISPA. Additionally, the ISPA provides the smaller or similar LQR cost than the GraSP as shown in Fig. \ref{fig:comparison_ISPA_GraSP} (b). From the simulation results, we demonstrate the ISPA can provide the better performance in the execution time than the GraSP. Additionally, it is noteworthy that the GraSP method is a specific sparsity-promoting method using $\ell_0$-ball, while the ISPA is a more general algorithm where various sparsity-promoting functions including $\ell_1$-ball and block $\ell_1$-ball can be used. 

\textcolor[rgb]{0,0,0}{\textbf{Remark:} In certain parameter settings, the GraSP exhibits faster speed than the ISPA. This can be attributed to the GraSP's utilization of Newton's method, a second-order method that provides a descent direction. In convex optimization problems, this method performs well and converges quickly. However, in the case of sparse LQR optimization, the problem is non-convex. While the algorithm can converge rapidly for some parameter values when the Hessian is well-calculated and positive definite, the presence of a negative definite or indefinite Hessian necessitates re-calculation with different points due to non-convexity. Consequently, the algorithm may exhibit slower convergence for some parameters.}

\subsection{Model-based deep neural network simulations}
We run simulations to check the performance of the model-based deep neural networks under the distributed system model introduced in Section \ref{subsec:simulation_prob} and a power system model. \textcolor[rgb]{0,0,0}{For the implementation of the DNN models, we use Tensorflow \cite{tensorflow2015-whitepaper} in version 1.12.}

We organize the DNNs based on the ISTA (DNN-ISTA), introduced in Fig.  \ref{fig:dataflow_ITA_RNN_BIGPIC} and the FISTA (DNN-FISTA) in Fig. \ref{fig:dataflow_FISTA_DNN_BIGPIC} with varying the number of layers $l$ from 10 to 50. Remark that the FISTA with the initial parameter $w_3^{(t)}=1$ is the same as the ISTA, and the DNNs based on ISTA or FISTA is the relaxation of the DNN based ISPA. For training the DNNs based on ISTA and FISTA, we calculate the loss value in \eqref{eq:training} and train the training parameters $w_1^{(t)}$, $w_2^{(t)}$, and $w_3^{(t)}$ via back-propagation in the DNNs having the simple shrinkage-threshold operator. With $t$ layers (correspondingly number of iterations in the ISTA), the average Normalized Mean Square Error (NMSE), denoted by $NMSE_t$, is calculated as follows:
\par\noindent\small
\begin{align}
	NMSE_t := \frac{1}{r}\sum_{i=1}^{r} \frac{\|  {{\bK}^{i}}^{(t)} - {\bK^{i}}^{\star} \|^2_F}{ \|{\bK^{i}}^{\star} \|^2_F},
\end{align}
\normalsize
where $r$ is the total number of data points in test dataset, and $\bK^{i}$ is the $i$-th feedback matrix corresponding to the $i$-th data point. In order to compare the performance of the DNNs with different layers to the ISTA, we perform the ISTA with the same number of iterations as the number of layers. \textcolor[rgb]{0,0,0}{More specifically, to obtain ${\bK^i}^{\star}$ for the $i$-th model, we executed ISTA for 10000 iterations without termination, considering it as the ground truth. Additionally, we ran ISTA again for only $t$ iterations to obtain ${\bK^i}^{(t)}$, enabling us to evaluate the gap between ${\bK^i}^{\star}$ and ${\bK^i}^{(t)}$. For DNN-ISTA and DNN-FISTA, we trained the model-based DNNs with $t$ layers. Subsequently, for each of the $r$ test data points, we obtained ${\bK^i}^{(t)}$ from DNN-ISTA and DNN-FISTA and calculated $NMSE_t$ with ${\bK^i}^{\star}$.}
 
\subsubsection{Distributed multi-agent control system model}
\label{subsubsec:Sim_distributed_model}
With the specific multi-agent system model in Section \ref{subsec:simulation_prob}, we generate 1100 dataset by adding noise following $\cN(0,0.1^2)$ to the non-zero elements of the state matrix $\bA$, and use 1000 dataset for training and 100 dataset for testing DNNs. In the generation of the dataset with the ISTA, we set other parameters as follows: $\gamma=1$, $\rho_{0}=100$, $\alpha=1.5$, and $\epsilon=10^{-4}$. \textcolor[rgb]{0,0,0}{We train the DNNs with 1000 epochs and 0.0001 step-size with the same initial parameters as ones that we used in the ISTA algorithm.}\footnote{\textcolor[rgb]{0,0,0}{For training DNN-ISTA with 30 layers, it took about five days on our Macbook pro 2016 with Intel Core i7 and 16 GB.}} Thus, the initial training parameters $w_1^{(t)}$, $w_2^{(t)}$, and $w_3^{(t)}$, $t=0,1,...,l$, are set to $1/100 (=1/\rho)$, $1/100(=\gamma/\rho)$,  and $1$ respectively.

Table \ref{tbl:dnn_error_N5} shows the test error in the average NMSE over 100 test dataset with different number of layers at $N=5$ multi-agents system model.  In the multi-agent system model, only with 30 numbers of layers (correspondingly 30 numbers of iterations in the ISTA), the DNN based on the FISTA achieves less than $10^{-3}$ test error in the NMSE from 100 test dataset, i.e., $r=100$. In order to have the similar test error, the ISTA requires 100 - 200 numbers of iterations. Therefore, around 70\% - 85\% of number of iterations (number of layers) can be reduced by using the data-driven approach. Since the computational complexity of the ISTA (or DNN) is proportional to the number of iterations (or number of layers), the computational complexity of the DNN can be reduced to the same amount as well with the optimally tuned weights obtained from the data-driven approach. In addition, from the comparison between the DNNs based on the ISTA and the FISTA, the DNN based on the FISTA improves 6.5\% in accuracy than the DNN based on ISTA with 30 layers.


\begin{table}[ht] 
\begin{center}
{\small
\begin{tabular}{|c|c|c|c|}
\hline
\# of layers, $t$ & ISTA 	& DNN-ISTA & DNN-FISTA \\ \hline\hline
10               &   6.839e-02  			& 	3.806e-03     		&    3.272e-03       \\ \hline
20			   &	 4.084e-02		 	&	1.178e-03	     	&   1.155e-03		 \\ \hline
30               &   2.480e-02   		&	6.503e-04			&   6.081e-04         \\ \hline
50               &   9.229e-03 			&  -        &    -        \\ \hline
100             &   1.047e-03   		&   -       &     -        \\ \hline
200             &    5.083e-05         &   -       &     -        \\ \hline
300             &    2.071e-06			&   -       &     -        \\ \hline
\end{tabular}
}
\caption{Test error in average $NMSE_t$ with $N=5$ multi-agent system model} \label{tbl:dnn_error_N5}
\end{center}
\end{table}



\subsubsection{16 machine power system model from power system toolbox}
In this simulation, we test our ISTA and its DNN implenetation in a power system model. We generate a small signal linearlized model from 16 machine system with 86 transmission lines and 68 buses by using Matlab Power System Toolbox (PST) \cite{chow1992toolbox}. In order to reduce the complexity of the system model, four generators in the system model are considered. We use the regularization term $G(\bK) = \| \bK \|_1$. With the parameter setting $(\gamma, \rho, MaxItr) = (1, 100, 500)$, we generate 100 data points for training and 10 data points for test by adding the noise $\cN(0,1)$ to the non-zero elements of the state matrix $\bA$. Other state matrices are set to $\bB=\bI$.  We calculate the average NMSE with the generated dataset from the ISTA and the estimate feedback matrix obtained from the DNNs by varying the number of layers (correspondingly the number of iterations in the ISTA). Table \ref{tbl:dnn_PST} shows the test error in the average NMSE. In the simulation, in order to achieve similar NMES, when the ISTA runs 400 iterations, the DNN based on FISTA only needs 20 layers, which reduces 95\% of iterations. 
\begin{table}[ht] 
\begin{center}
{\small
\begin{tabular}{|c|c|c|c|}
\hline
\# of layers, $t$ & ISTA 	& DNN-ISTA & DNN-FISTA \\ \hline\hline
10                &    4.425e-03  	    &  1.636e-03  &   2.602e-03 \\ \hline
20                &   4.229e-03         &   4.540e-04 	&   1.816e-04  \\ \hline
30                &   4.038e-03   		&   9.784e-05   &   9.651e-05  \\ \hline
40               &    3.852e-03         &    9.150e-05   &    7.654e-05  \\ \hline
50               &    3.671e-03 		     &    7.826e-05      &   5.379e-05   \\ \hline
100             &    2.842e-03 		&   -       &     -        \\ \hline
200             &    1.533e-03     	     &   -       &     -        \\ \hline
300             &    6.520e-04     	&   -       &     -        \\ \hline
400             &    1.558e-04     	&   -       &     -        \\ \hline
\end{tabular}
}
\caption{Test error in average $NMSE_t$ with PST 16 machine system model} \label{tbl:dnn_PST}
\end{center}
\end{table}

\subsection{Using a model-based deep neural network result as a warm-start of ISTA}
In this simulation, we consider a use-case where our model-based neural network is trained offline using a nominal system model. However, during runtime, if a more accurate system model is obtained, we employ ISTA to refine the feedback control. Essentially, we assume that we have a nominal system model denoted as $(\bA^{o}, \bB^{o})$, along with perturbed versions $\{ (\bA^{o} + \Delta \bA^i, \bB^{o} + \Delta \bB^{i}) \}_{i=1}^{r}$. By utilizing these perturbed versions as a dataset, we train DNN-ISTA offline, which allows us to obtain an approximate sparse controller. Subsequently, when we acquire a specific system model, we utilize the approximate sparse controller as a warm-start for ISTA to fine-tune it. 

For the nominal system model, we utilize the distributed multi-agent control system model described in Section \ref{subsec:simulation_prob}. DNN-ISTA with 30 layers in Section \ref{subsubsec:Sim_distributed_model} is employed to obtain an approximate sparse controller offline. To create a specific system model, we introduce randomly generated Gaussian noise following $\bN(0,0.1^2)$ to the non-zero elements of $\bA^{o}$ in the nominal system model. 

For comparison purposes, we execute ISTA as outlined in Algorithm \ref{alg:ISTA} using the specific system model. Additionally, we apply ISTA with the approximate sparse feedback matrix as a warm-start. To ensure statistical significance, we perform 100 trials with 100 different specific system models. The simulation results are summarized in Table \ref{tbl:ISTA_warmStart}. Notably, utilizing the warm-start approach results in a 63\% reduction in execution time and a 76\% decrease in the number of iterations with similar LQR cost and the sparsity level.

\begin{table}[ht] 
\begin{center}
{\small
\begin{tabular}{|p{11em}|c|c|}
\hline
Comparison	in average  											& ISTA 				& ISTA with warm-start \\ \hline\hline
Avg. execution time (sec.)              &    0.1845 	    &  0.0674   \\ \hline
Avg. number of iteration             &   370.20         &   88.84 	 \\ \hline
Avg. LQR cost               &   131.60   		&   131.57     \\ \hline
Avg. number of non-zero entries in $\bK^{\star}$             &    52.53         &    53.12 \\ \hline
\end{tabular}
}
\caption{\textcolor[rgb]{0,0,0}{Comparison result in average between ISTA and ISTA with warm-start}} \label{tbl:ISTA_warmStart}
\end{center}
\end{table}

\section{Conclusion and discussion}
\label{sec:conclusion}
In distributed control systems having multi-agents, the structure of a feedback matrix in the space state representation with feedback loop is related to communication links among agents. Since reducing the number of non-zeros in the off-diagonal of the feedback matrix can be understood as reducing the communication links, we consider to solve sparse LQR optimal control design problems for sparse feedback controllers. For that, we propose two algorithms - Iterative Shrinkage-Thresholding Algorithm (ISTA) and Iterative Sparse Projection Algorithm (ISPA). We show that our proposed algorithms can outperform the previous methods using the ADMM \cite{lin2013design} and the GraSP \cite{lian2017game}.  Moreover, in order to improve the performance of the proposed algorithms, we organize Deep Neural Network (DNN) models based on the proposed algorithms, where one layer of the DNNs represents the one iteration of the proposed algorithms. By using the data-driven approach, we can further reduce the number of iterations in the proposed algorithms, which can provide improved convergence speed. In our numerical experiments, we demonstrate that the DNNs can reduce 70\% - 95\% iterations in the algorithms.

\section*{Acknowledgment}
We thank Mihailo R. Jovanovic at USC for providing the simulation codes for the paper \cite{lin2013design}. And we also thank Feier Lian for providing the GraSP code used in \cite{lian2017game}, and Anaya Chakrabortty and Haoqi Ni at FREEDM center, NCSU for introducing this interesting problem and helping us generate the PST system model in the numerical experiments.



%
%

\bibliographystyle{IEEEbib}
\bibliography{refs_SparseK}
%
%

\end{document}